\documentclass[traditabstract]{aa}  

\usepackage{graphicx}
\usepackage{txfonts}
\usepackage{graphicx}
\usepackage{times}
\usepackage{natbib}
\usepackage{color}
\usepackage{multirow}
\usepackage{todonotes}
\usepackage{appendix}
\usepackage{array}
\usepackage{verbatim}

\newcommand{\enzo}{\it{\small ENZO}}

\begin{document} 

   \title{Proprieties of clumps and filaments around galaxy clusters}

   \author{M. Angelinelli\inst{1,2} \fnmsep\thanks{\email{matteo.angelinelli2@unibo.it}}
          \and
          S. Ettori\inst{2,3}
          \and
          F. Vazza\inst{1,4,5} 
          \and
          T.W. Jones\inst{6}
          }

   \institute{Dipartimento di Fisica e Astronomia, Università di Bologna, Via Gobetti 92/3, 40121, Bologna, Italy
         \and
             INAF, Osservatorio di Astrofisica e Scienza dello Spazio, via Pietro Gobetti 93/3, 40129 Bologna, Italy
         \and
             INFN, Sezione di Bologna, viale Berti Pichat 6/2, 40127 Bologna, Italy
         \and 
             Hamburger Sternwarte, University of Hamburg, Gojenbergsweg 112, 21029 Hamburg, Germany
         \and
            Istituto di Radio Astronomia, INAF, Via Gobetti 101, 40121 Bologna, Italy
        \and
            School of Physics and Astronomy, University of Minnesota, Minneapolis, MN, USA
             }

   \date{Received / Accepted}


\abstract    
   {We report on the possibility of studying the proprieties of cosmic diffuse baryons by studying self-gravitating clumps and filaments connected to galaxy clusters. While filaments are challenging to detect with X-ray observations, the higher density of clumps makes them visible and a viable tracer to study the thermodynamical proprieties of baryons undergoing accretion along cosmic web filaments onto galaxy clusters. We developed new algorithms to identify these structures and applied them to a set of non-radiative cosmological simulations of galaxy clusters at high resolution. We find that in those simulated clusters, the density and temperature of clumps are independent of the mass of the cluster where they reside. We detected a positive correlation between the filament temperature and the host cluster mass. The density and temperature of clumps and filaments also tended to correlate. Both the temperature and density decrease moving outward. We observed that clumps are hotter, more massive, and more luminous if identified closer to the cluster center. 
   Especially in the outermost cluster regions ($\sim$3$\cdot$R$_{500,c}$ or beyond), X-ray observations might already have the potential to locate cosmic filaments based on the distribution of clumps and to allow one to study the thermodynamics of diffuse baryons before they are processed by the intracluster medium.}


\maketitle

\section{Introduction} \label{sec:introduction}
On large scales, the matter in the Universe is organized into web-like patterns, the so-called cosmic web \citep{Bond96}. This structure is now mostly observed thanks to large optical surveys such as Sloan Digital Sky Survey \citep[SDSS,][]{Tegmark04}, 2 Micron All-Sky Survey \citep[2MASS,][]{Huchra12}, and VIMOS Public Extragalactic Redshift Survey \citep[VIPERS,][]{Guzzo14}, in which the large-scale distribution of the galaxies suggests the presence of filaments and knots. If the knots of the cosmic web are mainly associated with the galaxy clusters, the filaments are over-density regions which link different knots. The presence of these cosmic filaments is suggested by the spatial distribution of the galaxies; however, in recent years, diffuse gas filaments have also been mainly observed in the far-UV \citep{Nicastro18}, 
 by way of O VI absorption systems \citep{Danforth05,Tripp06} and in thermal soft X-rays \citep{2015Natur.528..105E}. 
 
 This gas is characterized by densities $\sim10^{-5} ~\rm cm^{-3}$ and temperatures from $10^5$ to $10^7 \ K$, values which led this gas phase to be defined as a warm-hot intergalactic medium (WHIM). However, a complete physical description of the WHIM proprieties is still unknown due to the low X-ray emissivity of the gas which occupies these filamentary regions. Cosmological simulations reproduce the cosmic web structure, enabling the study of the evolution and proprieties of the WHIM \citep[e.g.,][]{Cen99,dave01}. 
 Predictions by cosmological simulations suggest that $\sim$30-40\% of cosmic baryons are in the WHIM phase \citep{Martizzi19}, and the mass density associated with this phase might be efficiently estimated through absorption features in spectra obtained from background sources such as active galactic nuclei (AGN) or gamma-ray bursts \citep[GRB, e.g.,][]{Branchini09}.
 
Within cluster outskirts, baryon clumps and filaments have also been observed, although with some difficulty due to their low X-ray emissivities. For such cluster-embedded filaments, detection in the X-ray band requires sophisticated techniques such as deep, mosaiced observations \citep[see][for a detailed review on cluster outskirts]{walker19}. On the other hand, clumps have been found in the outskirts of the Coma cluster using Subaru weak-lensing mass maps by \citet{okabe14}, which identified sub-halos with masses around $10^{12} \ M_{\odot}$. Subsequently, using a sample of 31 galaxy clusters imaged by  ROSAT, \citet{eckert15}  were able to estimate the clumping factor by comparing the median and mean of the surface brightness profile. They found comparable results with Suzaku observations and theoretical predictions from previous numerical works \citep{2013MNRAS.432.3030R,vazza13}. 

Additionally, \citet{simionescu17} combined Suzaku X-ray observations and Sunyaev-Zel’dovich (SZ) effect maps derived from Planck telescope data for the Virgo cluster. They found an excess in the pressure profile obtained from the X-ray information alone that is explained by the presence of substantial gas density fluctuations in the cluster's outskirts (i.e., clumping). However, existing data do not yet provide a full characterization of these structures in cluster outskirts. As with the WHIM observations, the low X-ray brightness of clumps makes their detection very challenging. 

\begin{figure*}
\includegraphics[width=0.99\textwidth]{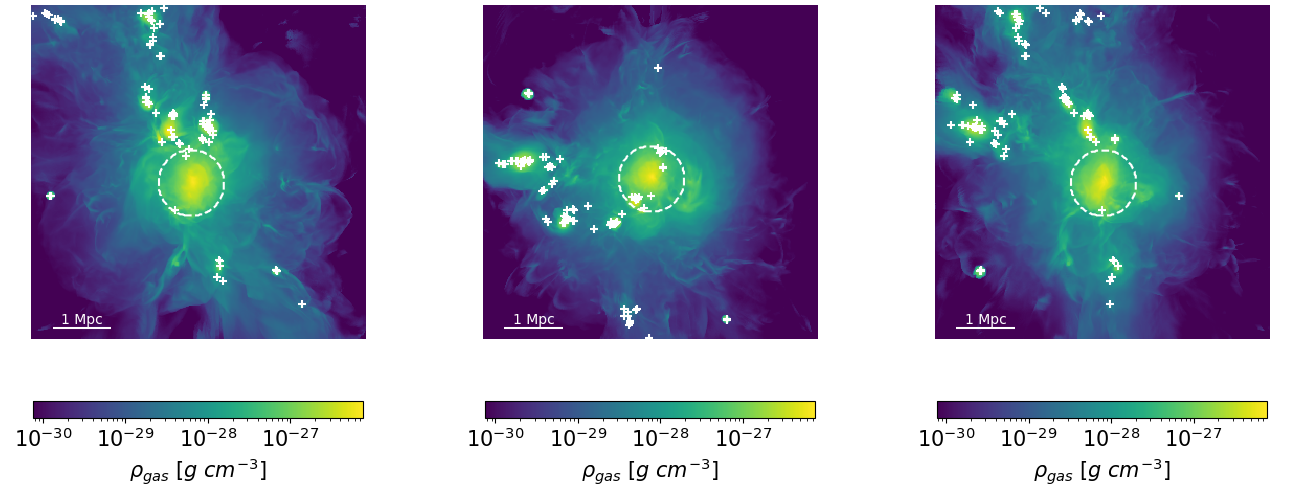}
\caption{Projected emission weighted gas density for the simulated cluster IT90\_2 a z=0.1, along three different lines of sight. The dashed circles represent R$_{\rm 500,c}$, while the white crosses are at the centers of the identified clumps.}
\label{fig:maps+clumpspos}
\end{figure*}

\begin{figure*}
\includegraphics[width=0.99\textwidth,height=0.7\textwidth]{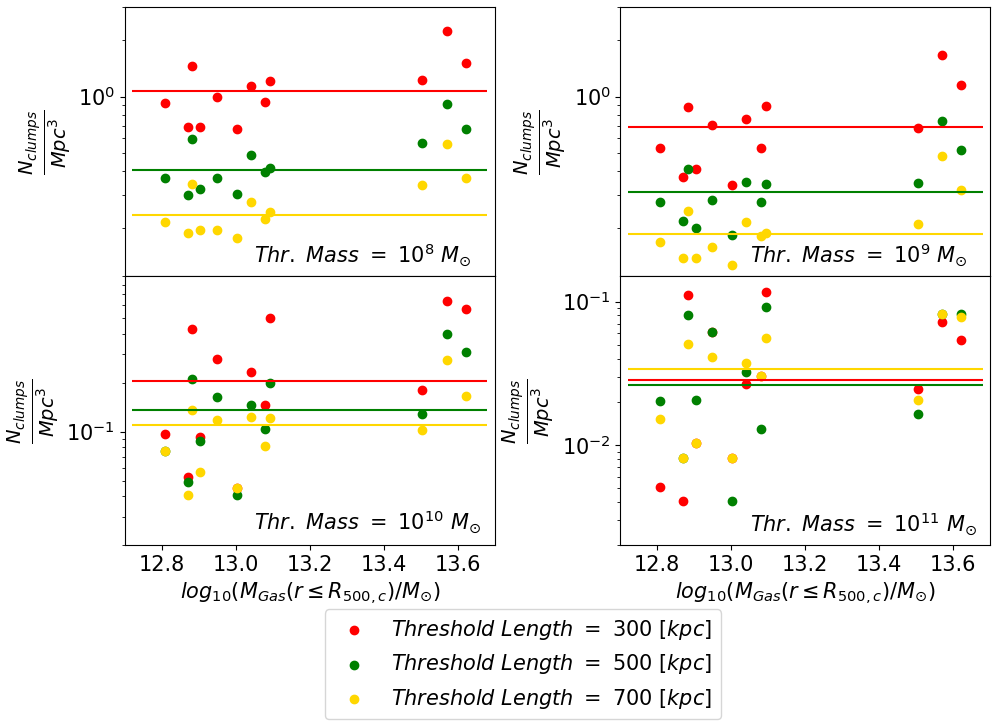}
\caption{Number of identified clumps per Mpc$^3$, as a function of the simulated host cluster's mass. The different panels show different threshold masses (top left: 10$^8$ M$_{\odot}$; top right: 10$^9$ M$_{\odot}$; bottom left: 10$^{10}$ M$_{\odot}$; bottom right: 10$^{11}$ M$_{\odot}$), while the different colors represent different threshold maximum distances (red: 300 kpc; green: 500 kpc; gold: 700 kpc). The solid lines, which have the same color-coding as the threshold maximum distances, represent the median values of clumps per Mpc$^3$ for the related threshold mass, computed on the whole simulated cluster sample.}
\label{fig:nclumps_clustermass}
\end{figure*}

Several numerical simulation studies have examined the impact of the clumps on the clusters' evolution \citep{nala11,vazza13,2013MNRAS.428.3274Z,2013MNRAS.432.3030R}. \citet{nala11} show that the presence of clumps in clusters' peripheries have to be taken into account to correctly reproduce the observed profiles of gas density and entropy derived from X-ray observations. In particular, they demonstrate that the average density as a function of radius is typically over-estimated if the clumps are not excised, which may lead to a spurious flattening of the gas entropy profile. 

\citet{2013MNRAS.428.3274Z} compare median values of density and pressure to the mean values of the same quantities.  
In their analysis, they find that the gas proprieties in radial shells are well reproduced by a log-normal PDF (Probability Density Function), with a tail. This tail corresponds to about $1\%$ of the matter in a given radial shell and effectively accounts for the gas inhomogeneities (clumps). With this approach the clumped component of the matter is easily separated from the diffuse counterpart.

\citet{2013MNRAS.432.3030R} study the relations between clumps and the measurement of the clusters' masses. They find that a correct description of the gas inhomogeneities is needed, because, otherwise, the reconstruction of the density profile introduces a bias in the estimate of the mass on the order of $\sim10\%$. They also introduce the residual clumpiness factor to describe the density inhomogeneities in the ICM. They find highly significant correlations between this parameter and the {\it y-}parameter profile derived from the SZ observations, but also with the azimuthal scatter in the X-ray surface brightness studies. They conclude that both {\it y-}parameter and azimuthal scatter are useful proxies for the residual clumpiness factor.

\citet{vazza13} study the relations between clumps and baryon fraction, clump detectability in some X-ray missions and also test the effects of cooling, feedback and numerical resolution on the estimates of clump proprieties. They find that the presence of clumps in the cluster's outskirts affects the derivation of the gas mass and this, in turn, affects the estimation of the baryon fraction. They show how in relaxed systems the introduced error by the gas clumps is $\sim10\%$, while it reaches $\sim20\%$ in disturbed systems. They produce mock X-ray maps in the soft X-ray band ($[0.5-2]$ keV) to investigate the detectability of the clumps. They find relations between the numbers of bright clumps and the dynamical state of the host cluster, but they do not find similar relations with a host cluster's mass. They also observe that the majority of the clumps are located in the radial range 0.6 $\le R/R_{200} \le$ 1.2, and conclude that for past and current X-ray missions (e.g., ROSAT, XMM-Newton, Chandra) disentangling diffuse from clumped emissions is difficult, owing to the required high brightness contrast. They also study the effects of cooling, feedback and numerical resolution on observable clump proprieties. We further address their conclusions in Sect.~\ref{sec:ITASCA}.

Filaments within clusters are even harder to isolate observationally than clumps. A growing set of works  do study the presence and the characterization of filaments between cluster pairs. A399-A401, A222-A223 and A3391-A3395  provide good examples \citep[][and reference therein]{Bonjean18,Werner08,Sugawara17}. 
Studies by \citet{2015Natur.528..105E,bulbul16} and \citet{connor19} analyze single filamentary structures around some individual clusters (Abell 2744, Abell 1750 and Abell 133, respectively). 

However, as noted in our introduction, multi-Mpc-scale filaments are major contributors to cosmic structure in general. So, we would expect to find their components in regions adjacent to individual clusters as well as within clusters where filamentary mass is actively accreting. For example, \citet{2015Natur.528..105E} observe filaments extending outside the viral radius of Abell 2744. They find a filamentary gas structure spanning $\sim8$ Mpc with a temperature $\sim10^{7}$K. They also estimate that the baryon fraction of this filaments is $\sim15\%$, and this allows them to conclude that a non-negligible part of the missing baryon mass in the Universe is in such cosmic filaments.
Recently, \citet{tanimura20} study a large sample of about 24,000 filaments identified in the SDSS survey. They select structures on scales from 30 to 100 Mpc. They used SZ information to exclude the contribution from clusters and groups and were able to identify and study the statistical signal of gas in filaments connecting pairs of halos.  In particular, they find temperatures $\sim10^6$ K, a typical overdensity $\delta$ $\sim20$, and a baryon fraction of about 8\%. 

Many numerical algorithms have been developed to study the complex hierarchy of structures in the simulated cosmic web \citep[e.g.,][for  reviews]{2014MNRAS.441.2923C,Libeskind18}. In general, while marking the knots of the cosmic web is a relatively simple task, isolating filaments involves more complex procedures, owing to the different possible definitions. Of specific relevance to the present work, we point out that available algorithms evidently do not reconstruct filaments in similar ways. However, it is also important to note that there still are important consistencies among those structure detecting algorithms. In particular, all the algorithms agree on the basic proprieties of cosmic web's voids and knots.   

For the rest of this paper, we focus on clumps and filaments within and in the proximity of clusters, as revealed in our simulations. The paper is organized as follows: in Sect.~\ref{sec:methods}, we present the simulations we used and the numerical algorithms we developed to identify clumps and filaments in the simulation data; in Sect.~\ref{sec:results}, we describe the resulting physical proprieties of clumps and filaments separately, and also the relations between them. We summarize the methods, the results and then discuss possible extensions of this work in Sect.~\ref{sec:conclusions}.

\begin{figure*}
\includegraphics[width=0.99\textwidth]{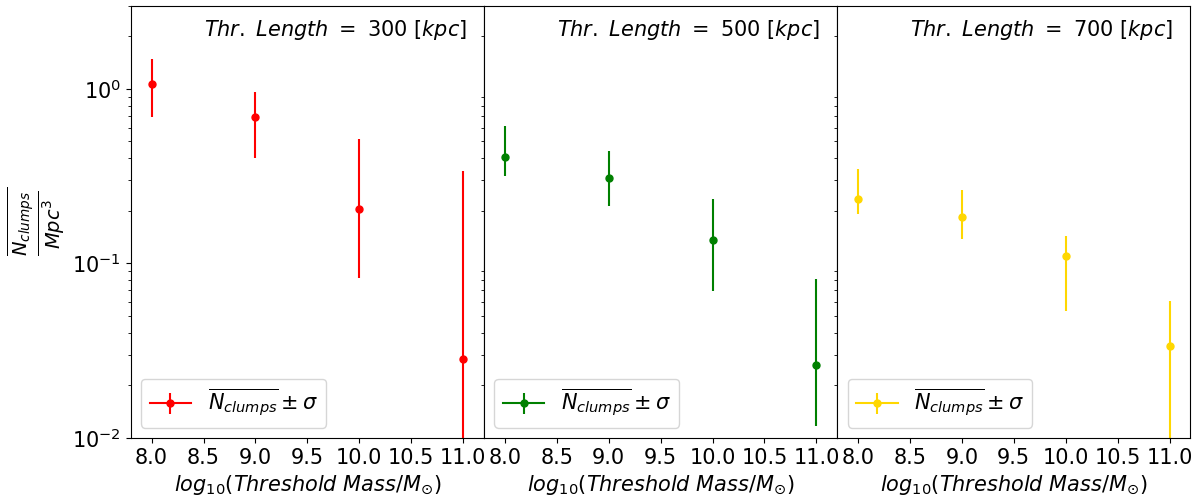}
\caption{Median number of identified clumps per Mpc$^3$, as a function of the threshold mass for the different maximum distance thresholds (left: 300 kpc; center: 500 kpc; right: 700 kpc). The error bars represent the 1$\sigma$ dispersion of the data.}
\label{fig:nclumps_thrmass}
\end{figure*}

\section{Methods} \label{sec:methods}

In this section, first we present the sample of cluster simulations used in our work, and then describe the numerical algorithms we develop to disentangle clumps from the diffuse gas in filaments. In the following analysis, we consider the baryon mass component  only; that is to say we  neglect  the  dark  matter  mass  contribution to the masses of clumps and filaments. This reflects the fact  that  our  clump  finder  algorithm  searches  for  baryon  density peaks, while the filament finder focuses on gas entropy, so it is also baryon-based. Considering that because collisional, baryonic matter and collisionless, dark matter do not obey exactly the same dynamics, there could be mismatches between the positions of peaks in the two  distributions.  This could in turn introduce  a  bias  in  the estimate  of  the  total  masses of  the clumps, which we want to minimize here. When and if necessary, we highlight different assumptions on the mass estimation.

\subsection{The Itasca simulated cluster sample}
\label{sec:ITASCA}
We use a subsample of the Itasca Simulated Clusters (ISC) sample for our analysis \footnote{http://cosmosimfrazza.myfreesites.net/isc-project. }. The sample used in this work is composed by set of 13 galaxy clusters in the $5\cdot 10^{13} \leq M_{100,c}/M_{\odot} \leq 4\cdot 10^{14}$ mass range (here, considering the combination of baryon and dark matter components) simulated at uniformly high spatial resolution with Adaptive Mesh Refinement and the Piecewise Parabolic method in the {\enzo} fluid dynamics code \citep[][]{enzo14}.  The ISC simulations do not include radiative processes and assume the WMAP7 $\Lambda$CDM cosmology \citep[][]{2011ApJS..192...18K}, with $\Omega_{\rm B} = 0.0445$, $\Omega_{\rm DM} = 0.2265$, $\Omega_{\Lambda} = 0.728$, Hubble parameter $h = 0.702$,  $\sigma_{8} = 0.8$ and a primordial index of $n=0.961$. Each cluster is generated from two levels of nested grids as initial conditions (each with $400^3$ cells and Dark Matter particles and covering $63^3 ~\rm Mpc^3$ and $31.5^3 \rm~Mpc^3$, respectively). At run time, we also impose two additional levels of static mesh refinement in a  $6.3^3 \rm ~Mpc^3$ box around each cluster, for a fixed $\Delta x = 19.6$ kpc/cell comoving resolution.  More information on the ISC sample can be found in \citet{va17turbo}, \citet{wi17b} and \citet{2018MNRAS.481L.120V}. 

These simulations do not include radiative gas cooling, nor the effect of heating from star forming regions, reionization or active galactic nuclei. As discuss in \citet{vazza13}, cooling, heating and numerical resolutions may affect the results of the analysis of the clump physics. In particular, those authors show that in numerical simulations which considered only cooling phenomena, the number and the density of the clumps are substantially different from non-radiative simulations, both for perturbed systems and relaxed ones. However, considering only cooling processes is not representative of the observed Universe \citep[e.g., clusters' cooling flow problem][]{Fabian12}. Numerical simulations in which AGN feedback is also taken into account, produce overdensities in agreement with the observed ones and in agreement also with the non-radiative simulations. In that context our work here is a first step intended to introduce a novel approach to this problem, while outlining proprieties found using these methods within the ensemble of simulated clusters we used to test it.

\begin{figure*}
\includegraphics[width=0.95\textwidth]{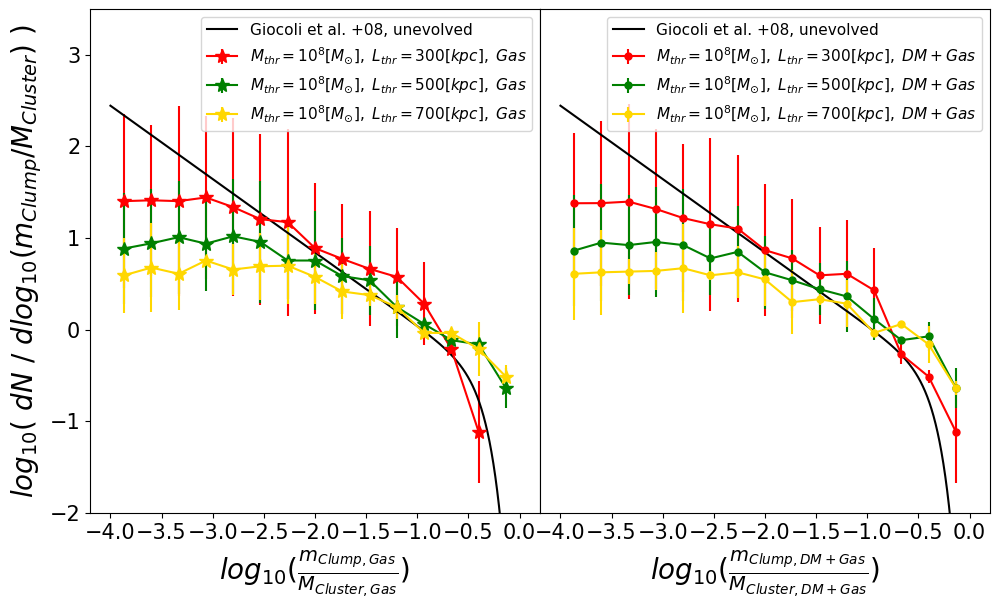}
\caption{Clump mass function for the $ \rm 10^{8} \ M_{\odot}$ mass threshold with different maximum distance thresholds (red: 300 kpc; green: 500 kpc; gold: 700 kpc. In the left panel we consider only the gas component of the mass, while on the right side we show the results for combined baryon and dark matter components. In both panels the black solid line represent the fitting formula proposed by \citet{giocoli08}.} 
\label{fig:clumps_massfunction}
\end{figure*}

\subsection{Identifying the clump structures}
\label{sec:clump}

In this analysis we consider clumps as the top 1\% of the densest regions within each radial shell from a given cluster center, following well-established evidence that clumps occupy the high density tail of the gas density distribution in the simulations \citep[e.g.,][]{2013MNRAS.428.3274Z}. 
Then, we add a second filter on the temperature of these structures. In the regions selected by the density threshold, we exclude all the cells with a temperature below 0.1 keV. This selection ensures us to identify regions which are comparable with the X-ray observations. To identify individual clumps, we have to implement a second, complementary algorithm which separates the initially identified region into specific clumps. This algorithm groups all the cells which are at a defined maximum distance from the initial reference cell. We also add a second criterion to exclude all the clumps which have a mass below a defined threshold mass. Figure~\ref{fig:maps+clumpspos} shows the results of this process, where the position of the clump's center (defined as the position of the maximum density in each clump) is identified for three different lines of sight of one of the clusters in our sample.

We test our algorithm with different values of maximum distance and mass threshold, and the results are reported in Fig.~\ref{fig:nclumps_clustermass}. It is apparent that increasing the host cluster's mass increases the number of identified clumps. This is observed for all the maximum distance thresholds when the threshold mass is below $\rm 10^{11} \ M_{\odot}$. Moreover, for these threshold masses, we also notice a decrease of the number of identified structures with an increase of the maximum distance threshold. Indeed, by increasing the maximum distance threshold, nearby structures, which are considered as different structures for small maximum distance threshold, are grouped into single, larger structures. For a fixed mass threshold, this behavior reduces the total number of identified structures with an increase of the maximum distance threshold. From the last panel of Fig.~\ref{fig:nclumps_clustermass}, we observe that none the relations between the number of identified clumps and the host cluster's mass or maximum distance thresholds are observed for the mass threshold  $\rm 10^{11} \ M_{\odot}$. This is, in fact, a consequence of the large magnitude of this threshold mass. Indeed, in some clusters in our sample, the algorithm does not detect clumps with masses exceeding $\rm 10^{11} \ M_{\odot}$, especially for smaller maximum distances. This suggests that, for these clusters, the clump threshold mass $\rm 10^{11} \ M_{\odot}$ is larger than any clumps that have formed.

In Fig.~\ref{fig:nclumps_thrmass}, we show the median number of identified clumps in our sample, as a function of the threshold mass. We notice that the number of identified clumps decreases when the threshold mass is increased for all the maximum distance thresholds. Moreover, as already observed for Fig.~\ref{fig:nclumps_clustermass}, the number of clumps decreases with increasing maximum distance threshold. Comparing Fig.~\ref{fig:nclumps_clustermass} and Fig.~\ref{fig:nclumps_thrmass}, we identify a threshold mass of $\rm 10^{8} \ M_{\odot}$ and a maximum distance threshold of 500 kpc as the best thresholds to identify clumps in our cluster sample. The chosen maximum distance threshold is in agreement with our previous work \citep{Angelinelli20}, where we studied turbulent motions in the same simulated clusters used here. In particular, we showed in that work how the peak of the Kolmogorov spectrum, which relates to the scale of the dominant energy-containing structures in the ICM, is on scales around 500 kpc. Those structures are, indeed, the same clumps studied in the current paper. Therefore, from the results presented in \citet{Angelinelli20} and the ones obtained from the algorithm presented above we can conclude that the maximum distance threshold of 500 kpc properly represents the typical upper scales of clumps in these clusters. Regarding the threshold mass, we assume $\rm 10^{8} \ M_{\odot}$ to avoid neglecting any appropriate clump structures. Indeed, if we compare the top panels of Fig.~\ref{fig:nclumps_clustermass} (which illustrate the results with a threshold mass of $\rm 10^{8} \ M_{\odot}$ and $\rm 10^{9} \ M_{\odot}$, respectively), we notice that the number of clumps in a Mpc$^3$ volume is almost the same. When we increase the threshold mass (bottom panels of Fig.~\ref{fig:nclumps_clustermass}), the number density of clumps drops quickly, suggesting that most of the clumps have masses lower than $10^{10} \ M_{\odot}$. Therefore, to sample properly the clump population in these clusters, we adopt the minimum threshold mass of $10^{8} \ M_{\odot}$.

Using these thresholds, we compare the mass function of the identified clumps with the results presented by \citet{giocoli08}. In that work, the authors studied a sample of N-body simulated dark-matter-only galaxy clusters. They focused on the mass-loss rate and the related mass function of the sub-halos present in their simulation boxes. They demonstrated that the mass function of sub-halos is universal, and they described it with the following fitting formula: 
\begin{equation}
    \rm \frac{dN}{d ln (m_{v}/M_{0})} = N_{0} \ x^{-\alpha} \ e^{-6.283 \cdot x^{3}},  \qquad x=\biggl| \frac{m_v}{\alpha \cdot M_{0}} \biggl|
\end{equation}
where $\rm m_v$ is the clumps' mass and $\rm M_0$ is the cluster's mass, while $\alpha$=0.8 and N$_0$=0.21. We compare this fitting formula with our mass distributions. The results are plotted in Fig.~\ref{fig:clumps_massfunction}.  

Our results including both baryons and dark matter are consistent with the fitting formula presented by \citet{giocoli08} (right panel of Fig.~\ref{fig:clumps_massfunction}). However, some differences in the distribution at low masses appear because of different mass resolutions in the two studies. When we consider only the gas component of our simulations, the mass distribution is well-described by the fitting formula proposed by \citet{giocoli08}. This consistent behavior is explained by a relative baryon component of clump mass that is almost the same as for the central, host cluster. Indeed, as this would suggest, the curves in the left and the right panel of Fig.~\ref{fig:clumps_massfunction} have similar trends. 

\begin{figure}
\includegraphics[width=0.49\textwidth]{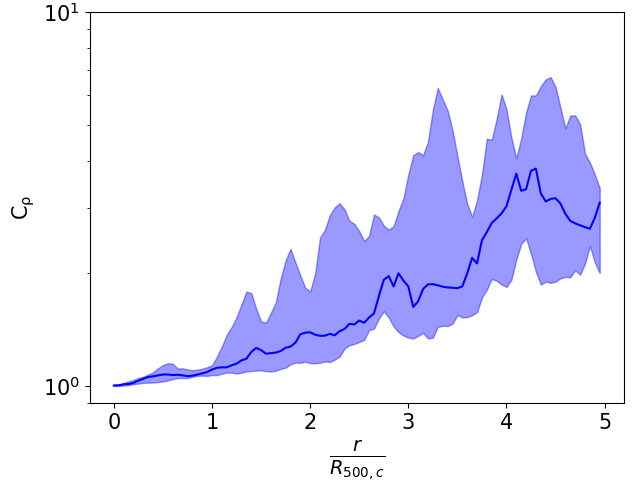}
\caption{Radial profile of clumping factor for the whole clusters sample. The blue solid line represents the median value, while the shadow region is enclosed between the 16$^{th}$ and the 84$^{th}$ percentile of the distribution at any radii.}
\label{fig:clumpingfactor_clus}
\end{figure}
\begin{figure}

\includegraphics[width=0.49\textwidth]{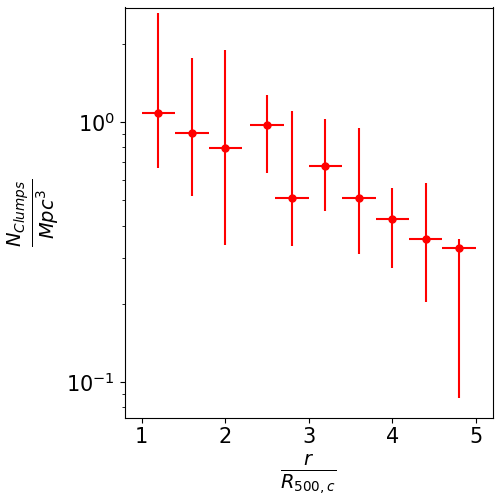}
\caption{Number density of clumps computed on ten radial bins between R$_{500,c}$ and 5$\cdot$R$_{500,c}$. The red dots are the median clumps' number density in each bin. The x-axis errors represent the length of radial bins, while the y-axis ones represent the 16$^{th}$ and 84$^{th}$ percentiles of the bin's number density distribution.}
\label{fig:nclumps_radial}
\end{figure}

As a next step in this analysis we compute the clumping factor for our full cluster sample. We define the clumping factor at each radius $r$ as: 
\begin{equation}
C_{\rho}(r) = \sqrt{\frac{\langle \rho(r)^2 \rangle}{\langle \rho(r) \rangle^2}}, 
\label{eq:clumping}    
\end{equation}
where the means are computed within a shell with radius $r$ \citep[e.g.,][]{nala11}. 
 In Fig.~\ref{fig:clumpingfactor_clus}, we show the median value of $C_{\rho}$, along with the 16$^{th}$ and 84$^{th}$ percentiles, at each radius. Similar to \citet{vazza13}, we observe an increment of clumping factor with the radial distance, and measure values which are in good agreement with the ones estimated in ,for example, \citet{nala11} and \citet{vazza13}. 

In Fig.~\ref{fig:nclumps_radial} we also present the relation obtained for our clusters between the clump number density and the radial distance from cluster center. We adopt the distance and mass thresholds described above and we use ten equally spaced radial shells from R$_{500,c}$ up to 5$\cdot$R$_{500,c}$. 
It is clear that the clump number density decreases with the radial distance, that is by a factor $\simeq$3 from the inner to the outer bin.
Comparing Fig.~\ref{fig:clumpingfactor_clus} and Fig.~\ref{fig:nclumps_radial} we conclude that, even though the clumping factor increases with radius, the clump number density decreases. These different behaviors suggest that the observed decreased number of clumps with the radial distance is not an artifact of our clump finder algorithm, but, rather, is a true property of the clump population. 
To further test the dependence of our results on the numerical resolution, 
in Appendix \ref{app:testres} we computed the clumping factor, as well as the clumps number density, for four different re-simulations of the same cluster, at increasing spatial resolution. Briefly, we find that the Itasca cluster simulation sample numerical resolution ($\sim$19.6 comoving kpc) is sufficient to study the  structures which our clump finder algorithm identifies. That is, our detected trends are not found to depend significantly on resolution, for $\Delta x \leq 32 \rm kpc$.

\subsection{The ${\rm \frac{V_{rad}^2}{K}}$ estimator}
\label{sec:vradk}

\begin{figure*}
\includegraphics[width=0.99\textwidth]{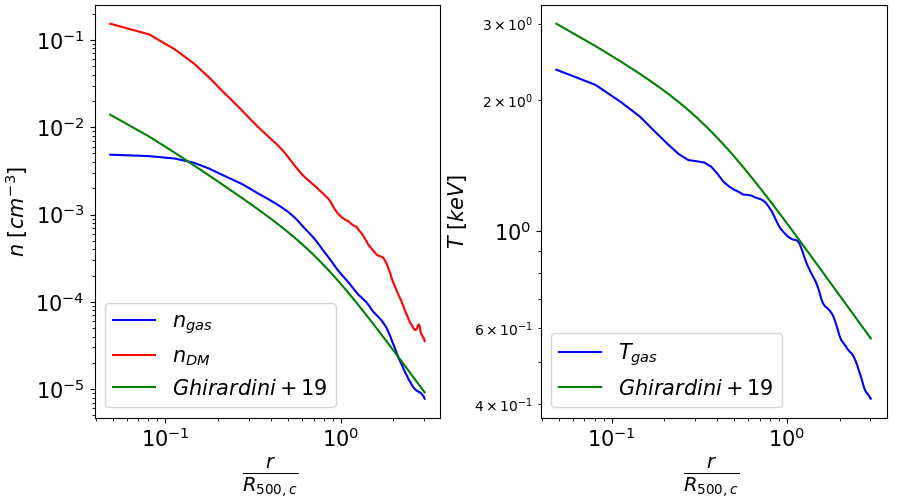}
\caption{Comparison between density and temperature profiles of the simulated cluster IT90\_2 ($M_{Tot,500,c} = 6.68 \cdot 10^{13} M_{\odot}, R_{500,c} =$ 555 kpc, $z=0.1$) and the universal profile proposed by \citet{Ghirardini:2019}. (Left) Gas density (blue), dark matter density (red) and universal (green) profiles in cm$^{-3}$ units. (Right) Gas temperature (blue) and universal (green) profiles in keV units. In both panels the profiles are computed from the cluster's center out to $3 \cdot R_{ 500,c}$}
\label{fig:IT90_2_profile}
\end{figure*}

\begin{figure*}
\includegraphics[width=0.99\textwidth]{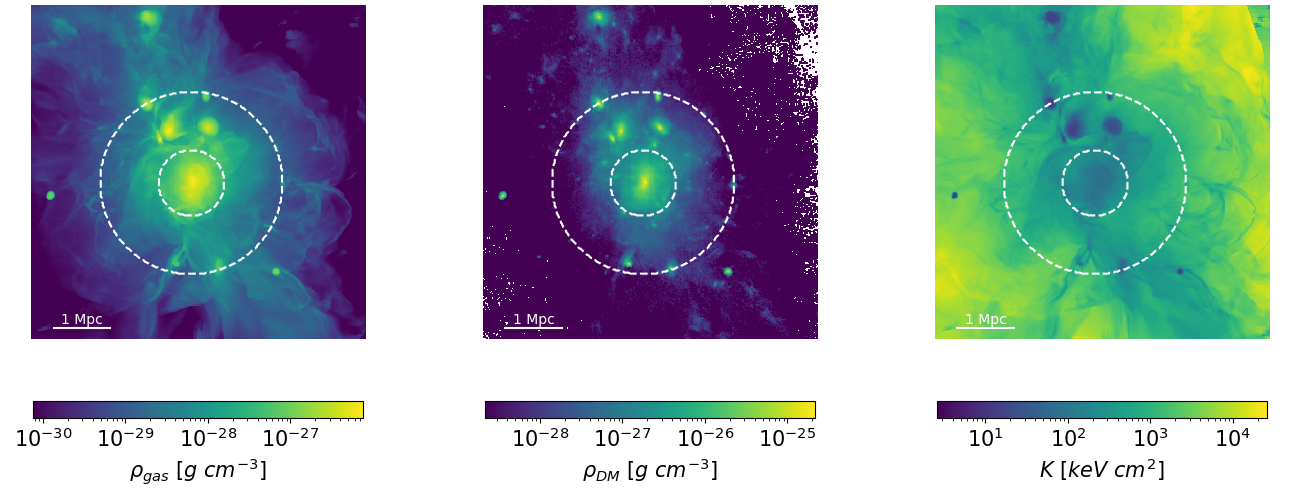}
\caption{Projected emission weighted gas density (left panel), dark matter density (central panel) and entropy (right panel) for the cluster IT90\_2 at $z=0.1$. The dashed circles represent $R_{\rm 500,c}$ and $2.8 \cdot R_{\rm 500,c}$.}
\label{fig:IT90_2_maps}
\end{figure*}

In this section, we use the simulated cluster IT90\_2 as an example to explore a novel {\it{filament}} identifier tool.  IT90\_2 is described at $z = 0.1$ by density and temperature radial profiles shown in Fig.~\ref{fig:IT90_2_profile}. These profiles follow well the rescaled universal cluster profiles proposed by \citet{Ghirardini:2019}.
The projected IT90\_2 gas density and dark matter density maps at $z = 0.1$ are shown in Fig.~\ref{fig:IT90_2_maps}, together with the projected entropy map.

To identify the diffuse gas in the filaments within and adjacent to the clusters, we define a proxy based on gas entropy and gas radial velocity. Indeed, we expect that these structures, having ``fallen'' out of the web, are moving toward the cluster center mainly in the radial direction, and with a velocity determined by gravity, comparable with the local in-fall velocity. Furthermore, the absence of a strong X-ray detection of filaments in real clusters so far suggests that their densities and temperatures are quite different from clump densities and temperatures. Specifically, the filaments are expected to be relatively cooler and very likely with low entropy. In our entropy-radial-velocity-based filament finding algorithm we define gas entropy, $K(i)$ in volume element $i$, as
\begin{equation}
\rm K(i) = T(i) \cdot n(i)^{-2/3}    
\end{equation}
and the radial velocity, $V_{rad}(i)$, as the radial projection of the velocity field, relative to the cluster center. The center was defined as the position of the maximum of the thermal energy density of the gas.
 
Now we consider the ratio between the $V_{rad}^2$ and $K$.
In the self-similar cluster formation approximation, the ratio of entropy to free-fall velocity is independent of the host cluster mass:
\begin{equation}
\rm \frac{V_{rad}^2}{K} \propto \frac{M^{2/3}}{M^{2/3}} = CONST.  
\end{equation}

\begin{figure}
\includegraphics[width=0.49\textwidth]{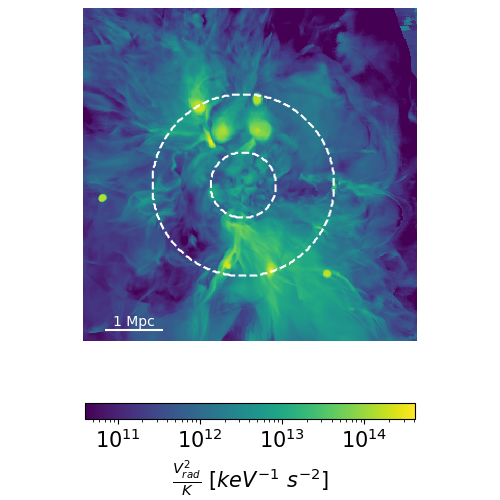}
\caption{Projected emission weighted $V_{rad}^2/K$ for the cluster IT90\_2 at z=0.1. The dashed circles represent R$_{\rm 500,c}$ and 2.8$\cdot$R$_{\rm 500,c}$.}
\label{fig:IT90_2_vrad}
\end{figure}

In Fig.~\ref{fig:IT90_2_vrad}, we show the projected emission weighted map of the ratio ${\rm \frac{V_{rad}^2}{K}}$. We note that filaments show higher values of this ratio than those of their ICM surroundings, especially reflecting reduced filament entropy. Utilizing this bias toward large $V_{rad}^2/K$, we include as filaments the regions in each radial shell {\it j}, with 
\begin{equation} \label{eq:filam}
\rm    \frac{V_{infall}^{2}(j)}{0.05\cdot\overline{K(j)}} \ < \ \frac{V_{rad}^{2}(j)}{K(j)} \ < \ \frac{V_{infall}^{2}(j)}{0.01\cdot\overline{K(j)}}
\end{equation}
where $\rm V_{infall}^{2}(j)$ is the free-in-fall velocity in the radial shell {\it j} given by
\begin{equation}
 \rm   V_{infall}^{2}(j) = \frac{G \cdot M(<j)}{r(j)}
\end{equation}
and $\rm \overline{K(j)}$ is the mean entropy in the radial shell {\it j}. As noted our selection is mainly based upon the gas entropy, because clumps and filaments have comparable radial velocities close to the free-in-fall velocity. But, they do differ by their entropy. The limits of 1\% and 5\% are imposed to exclude clumps and diffuse gas from the filament analysis. Indeed, as observed from Fig.~\ref{fig:IT90_2_vrad}, filamentary structures are described by high values of $V_{rad}^2/K$, but the highest values are actually reached in the presence of clumps. Therefore, we exclude the regions with the highest values of $V_{rad}^2/K$ in order to avoid the presence of clumps (Fig.~\ref{fig:IT90_2_vrad}). But, we need also a lower limit to select only filaments and to exclude the diffuse gas. 

\begin{figure*}
\includegraphics[width=0.99\textwidth]{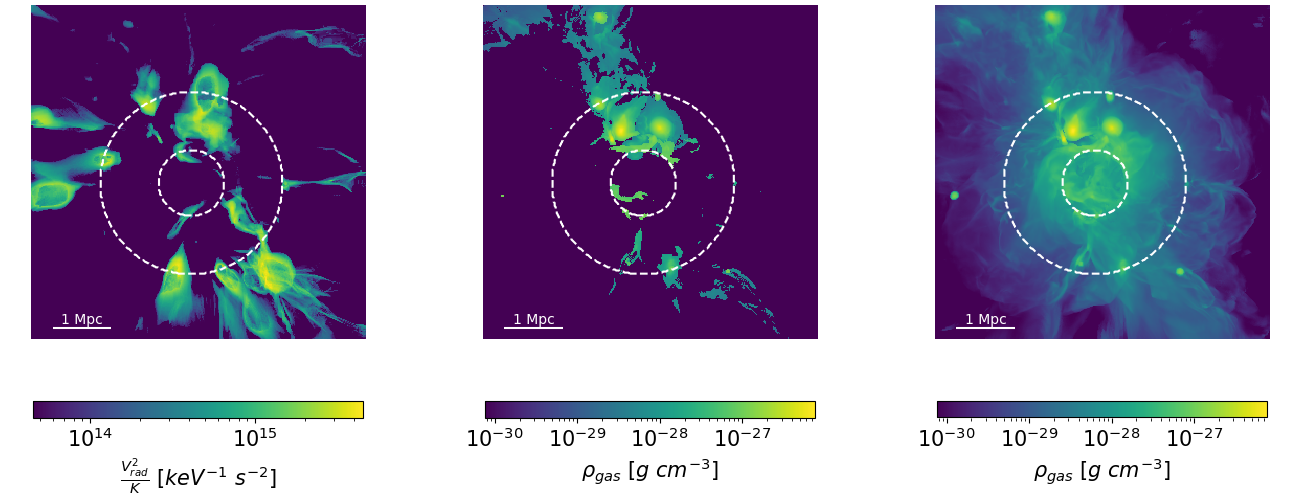}
\caption{Visual comparison between filaments, clumps, and WHIM. Left panel: projected integrated $V_{rad}^2/K$ for the filaments; center panel: projected emission weighted density for the clumps; right panel: projected emission weighted density for the gas with temperature between $10^{5}$K and $10^{7}$K. The cluster used is IT90\_2 at z=0.1. The dashed circles represent R$_{\rm 500,c}$ and 2.8$\cdot$R$_{\rm 500,c}$.}
\label{fig:IT90_2_filam_clump}
\end{figure*}

We also add a filter on the density of the gas. We consider gas to be part of a filament only when it has a density greater than the critical density of the Universe $\rm \rho_{c}(z)$, in order to avoid selecting under-dense regions, especially in the cluster outskirts. In Fig.~\ref{fig:IT90_2_filam_clump} we compare the map of the clumpy region against the filamentary structures map. We also add a third panel in which we display regions selected using the standard definition of WHIM ($10^5$ K$\le$T$_{gas}\le 10^7$ K ). We note that our definitions of clumps, derived by the clump finder algorithm, and filaments, based on the $V_{rad}^2/K$ proxy, select regions distinct from the standard definition of WHIM gas, which is based only on a temperature criterion.
In fact, looking at the right-most panel in Fig.~\ref{fig:IT90_2_filam_clump}, we observe how the standard definition of WHIM highlights relatively diffuse regions not directly associated with clumps or filaments. 

\begin{figure*}
\includegraphics[width=0.99\textwidth,height=0.4\textwidth]{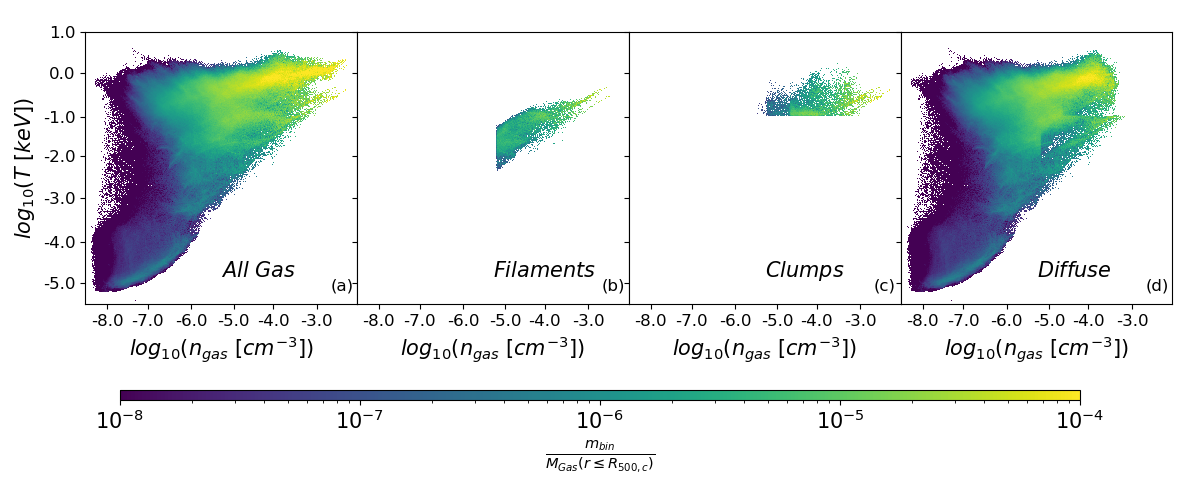}
\caption{Phase diagrams for the cluster IT90\_2 at z=0.1. (a): all gas present in the simulated box enclosing IT90\_2; (b): filaments selected by the $V_{rad}^2/K$ proxy; (c): clumps identified as described in Sect.~\ref{sec:clump}; (d): Diffuse gas located on radii over R$_{500,c}$ and not organized into clumps or filaments. The color-coding identifies the mass enclosed in each single bin, normalized by the central cluster's gas mass.}
\label{fig:IT90_2_phase}
\end{figure*}

In Fig.~\ref{fig:IT90_2_phase}, we present a gas-phase diagram analysis for the cluster IT90\_2 at $z=0.1$ in order to characterize these distinct distributions. We observe how clumps and filaments occupy different regions in the phase diagram. Indeed, filament temperatures are mainly below 0.1 keV, while clump temperatures reach peaks greater than 1 keV. However, the density distributions are actually quite similar, with values between $10^{-5}$ cm$^{-3}$ and $10^{-2}$ cm$^{-3}$. These differences assure us that our selection algorithms do not confuse the phases. In addition, we notice that the clumps seem to lie physically at the tips of the filaments pointing to related origins for these two phases. We study this finding in more in detail in Sect.~\ref{sec:physicalclumpfilam}. Moreover, the last panel of Fig.~\ref{fig:IT90_2_phase} shows the phase diagram for the gas component which we called "Diffuse". This gas is located at radii beyond R$_{500,c}$ and it is not organized into clumps or filaments.  

\section{Results} \label{sec:results}

In the following, we refer to "clumps" when we present results associated with the cells selected by the clump finder algorithm. Otherwise, when we refer to cells selected by the $V_{rad}^2/K$ estimator, we use the term "filaments". In Sect.~\ref{sec:radialtrends}, we compare the results obtained in two different radial shells of our clusters. As mentioned above, we define the center of a cluster as the position of the peak in the thermal energy of the gas. This definition of the center provides the most stable identification criterion, including highly perturbed systems. As outlined above, we focus on cluster peripheries. Hence, the inner shell considers all the radii enclosed in $R_{500,c} \leq r \leq 2.8 \cdot R_{500,c}$, whereas the outer shell enclose radii between $2.8 \cdot R_{500,c} \leq r \leq 5 \cdot R_{500,c}$. This radial analysis allows us to better assess the dependence of physical proprieties on distance from the cluster center. 

\begin{figure*}
\includegraphics[width=0.99\textwidth]{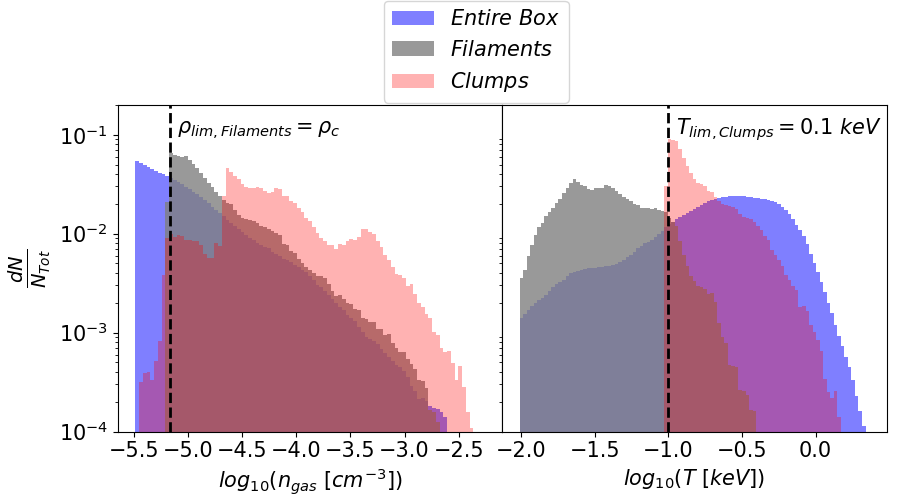}
\caption{Density and temperature distributions for the different selected regions for the IT90\_2 cluster at z=0.1: the blue areas represent the entire contents of the simulated box, gray areas characterize regions identified as filaments and red areas characterize clumps. Each histogram is normalized to the total number of cells present for the specified selection. The black dashed line in the left panel represents the $\rho_c(z)$, used as lower limit for the density in the $V_{rad}^2/K$ procedure. On the right panel the black dashed line is the lower temperature limit of 0.1 keV, adopted for the clump finder algorithm as explained in the text.}

\label{fig:IT90_2_nhistothisto}
\end{figure*}

Unless specified otherwise, we shall use the median, along with 16$^{th}$ and 84$^{th}$ distribution percentile boundaries to quote the results of our statistical analysis. These measures are justified by the distribution forms of physical proprieties we are studying. An example is given in Fig.~\ref{fig:IT90_2_nhistothisto} for the cluster IT90\_2. We notice that the distributions of both density and temperature are far from being described by a symmetrical Gaussians, making the mean and standard deviation values weaker representative measures than the median and the two above percentiles to describe  these distributions. 

\subsection{Physical proprieties of clumps and filaments}
\label{sec:physicalclumpfilam}

\begin{figure*}
\includegraphics[width=0.99\textwidth]{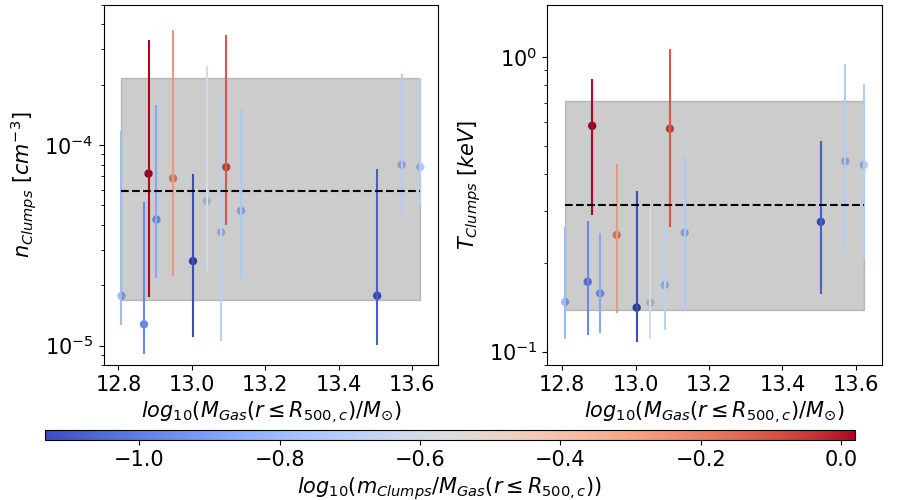}
\caption{Clump density (on the left, in cm$^{-3}$) and temperature (on the right, in keV) as a function of M$_{\rm 500,c}$ cluster gas mass. In both the panels, the dots represent the median value, while the error bars are the 16$^{th}$ and 84$^{th}$ distribution percentiles. The color-coding is the same for both the panels and it identifies the ratio between the total clump mass and the M$_{\rm 500,c}$ cluster mass. The black dashed lines represent the median values of density and temperature for the whole clump population. The shadow gray regions are enclosed in the 16$^{th}$ and 84$^{th}$ of the density and temperature distributions.}
\label{fig:n_t_clumps}
\end{figure*}

We now consider density and temperature of the cells selected by the clump finder algorithm. For each cluster we compute the median values of the density and temperature distributions and identify the locations of their 16$^{th}$ and 84$^{th}$ percentiles. We also determined masses of individual clumps. 
In Fig.~\ref{fig:n_t_clumps}, we show the median values of density and temperature as functions of cluster mass. We notice an almost flat distribution of the density against cluster mass around a value of $0.6^{+1.6}_{-0.4}\cdot 10^{-4}$ cm$^{-3}$.

We also observe that relatively few objects are far from this value. From the color-coding of Fig.~\ref{fig:n_t_clumps}, we can see that those objects have larger ratios of clump mass over cluster mass. In fact they are merging systems, for which our clump finder algorithm can consider the  baryons in the less massive interacting object as a mass clump  rather than a component of a merging halo. This will bias the total clump baryon mass estimation for those clusters.

The relation between cluster mass and clump temperature is shown in Fig.~\ref{fig:n_t_clumps} (right panel). The clump temperature distribution can be characterized as $0.3^{+0.4}_{-0.2}$ keV. We note that the median value $\approx$ 0.3 keV, while a factor three larger than the 0.1 keV lower temperature limit we adopted in our clump identification algorithm (see Sect.~\ref{sec:clump}), is still quite small. So, it is clear that any X-ray detection of clumps, although possible, will be very challenging. As already discussed for the density distribution, now for the temperature distribution, we have some clumps with median values far from the median of the full clump population with similar masses. We believe these outliers represent merging objects.

\begin{figure}
\includegraphics[width=0.48\textwidth]{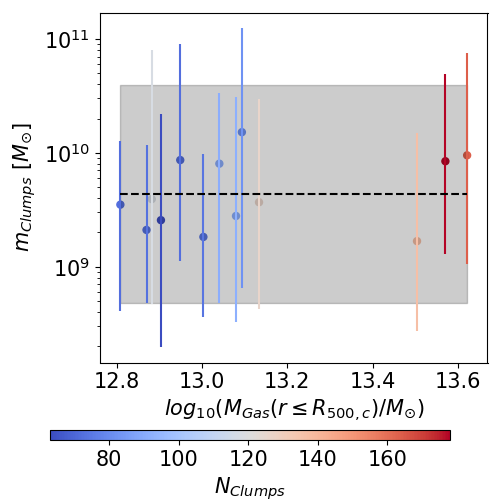}
\caption{Clump mass as a function of central cluster mass. The dots represent the median value, while the error bars are the 16$^{th}$ and 84$^{th}$ distribution percentiles. The color-coding identifies the total number of clumps found by the clump finder algorithm. The black dashed line represents the median value computed for the entire clump population. The gray shadow region is enclosed between the 16$^{th}$ and 84$^{th}$ percentiles of the clump mass distribution.}
\label{fig:m_clumps}
\end{figure}

\begin{figure}
\includegraphics[width=0.48\textwidth]{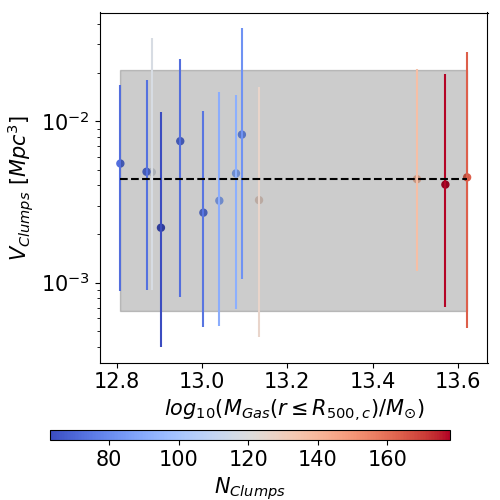}
\caption{Clump volume as a function of central cluster mass. The dots represent the median value, while the error bars are the 16$^{th}$ and 84$^{th}$ distribution percentiles. The color-coding identifies the total number of clumps identified by the clump finder algorithm. The black dashed line represents the median value computed for the entire clump population. The gray shadow region is enclosed between the 16$^{th}$ and 84$^{th}$ percentiles of the clump volume distribution.}
\label{fig:v_clumps}
\end{figure}

For each identified clump, we compute both mass and volume. We study the distributions of these quantities also in relation to the cluster's mass. The results are shown in Fig.~\ref{fig:m_clumps} for the masses and in Fig.~\ref{fig:v_clumps} for the volumes. For both distributions, we obtain median values quite independent of the host cluster's mass. For the clump masses, the median value is $0.44^{+3.53}_{-0.38}\cdot 10^{10}$ M$_{\odot}$, while for the volumes it is $0.4^{+1.6}_{-0.3}\cdot 10^{-2}$ Mpc$^{3}$. In the following, using the $V_{rad}^2 / K$ criterion introduced above, we investigate the relations between proprieties of clumps and those of the surrounding medium.

\begin{figure*}
\includegraphics[width=0.99\textwidth]{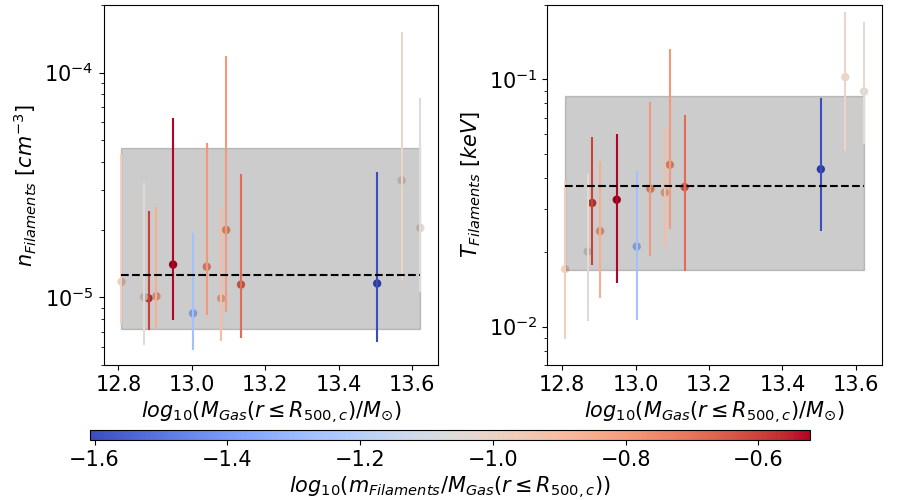}
\caption{Filament density (on the left, in cm$^{-3}$ units) and Filament temperature (on the right, in keV units) as functions of M$_{\rm 500,c}$ cluster  {\it{gas}} mass. In both panels, the dots represent the median value, while the error bars are the 16$^{th}$ and 84$^{th}$ distribution percentiles. The color-coding is the same for both the panels. It identifies the ratio between the total filament gas mass and the M$_{\rm 500,c}$ cluster gas mass. The black dashed lines represent the median population values of density and temperature for the whole filament population. The shadow gray regions are enclosed in the 16$^{th}$ and 84$^{th}$ of the density and temperature distributions.}
\label{fig:n_t_filam}
\end{figure*}

For both clumps and filaments, we analyze the relations between the density and the temperature of the selected regions, including their dependence on the host cluster's mass. The results of these analyzes are shown in Fig.~\ref{fig:n_t_filam}. The density distribution appears to be independent of cluster mass (left panel of Fig.~\ref{fig:n_t_filam}) with a median value of $1.3^{+3.4}_{-0.5}\cdot 10^{-5}$ cm$^{-3}$.

On the other hand, the temperature of the filaments seems to increase slightly with the cluster's mass (with a Pearson's correlation index $\rho_{XY}=0.88$;  see App.~\ref{app:pearson} for details) around a median value of $0.04^{+0.05}_{-0.02}$ keV (right panel of Fig.~\ref{fig:n_t_filam}). This value is less than 0.1 keV, the lower limit we adopt for X-ray detection, confirming that the observations of X-ray emitting filaments are on average very challenging with current (and even future) X-ray telescopes. These behaviors also are consistent with the fact that the few significant observational X-ray confirmations that exist are typically associated with massive cluster systems.

The color-coding of Fig.~\ref{fig:n_t_filam} identifies the ratio between filament masses and cluster masses. Distinct from the clump situation, our selection is not able to separate filaments as single, isolated objects, so we need to consider the total gas mass of all filaments connected to a given cluster. 
By comparing Fig.~\ref{fig:n_t_clumps} and Fig.~\ref{fig:n_t_filam}, we observe that some typical values can characterize the entire populations of clumps and filaments. Moreover, we do not find strong correlation between clump or filament proprieties with the host cluster mass, excepting the  filament temperature relation mentioned above. Even so, for these quantities we could reasonably describe the full filament population using single values. 

\begin{figure*}
\includegraphics[width=0.99\textwidth]{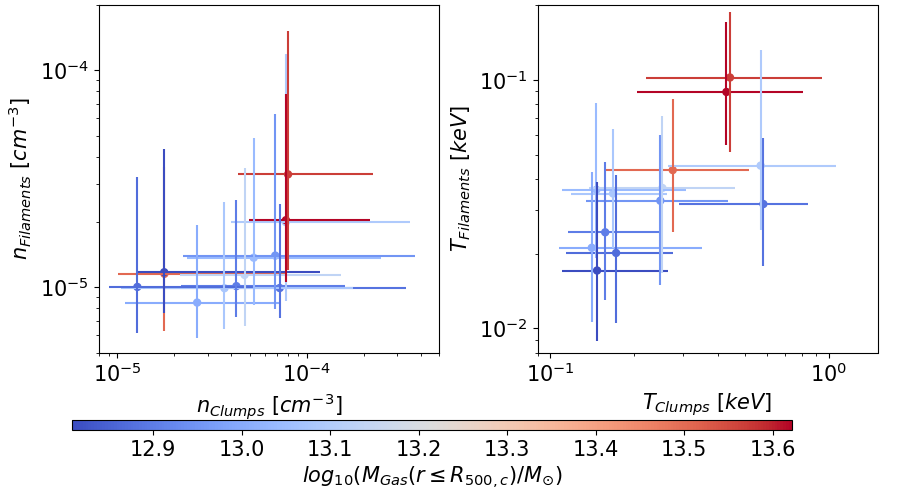}
\caption{Comparison of density (in cm$^{-3}$ units) and temperature (in keV units) for clumps (on the x-axis) and filaments (on the y-axis), for each single cluster. The dots are the median values, while the error bars are the 16$^{th}$ and 84$^{th}$ percentiles on both the axes. The color-coding is the same for both the panels and it identifies the M$_{\rm 500,c}$ cluster gas mass.} 
\label{fig:n_t_clumpsvsfilam}
\end{figure*}

Up until this point we have concentrated on issues specific to either clumps or filaments. We now briefly compare the relative densities and temperatures for clumps and filaments populations of each single cluster. In Fig.~\ref{fig:n_t_clumpsvsfilam}, we present the results of these analyzes. Both for the density and the temperature, there appear to be moderate correlations between clump and filament proprieties. For the density comparison between clumps and filaments we measure a Pearson's index, $\rho_{XY}=0.66$, while for  temperature, we estimate a somewhat weaker correlation with $\rho_{XY}=0.54$ (see App.~\ref{app:pearson}).

\subsection{Analysis of the radial trends}
\label{sec:radialtrends}

\begin{figure}
\includegraphics[width=0.49\textwidth]{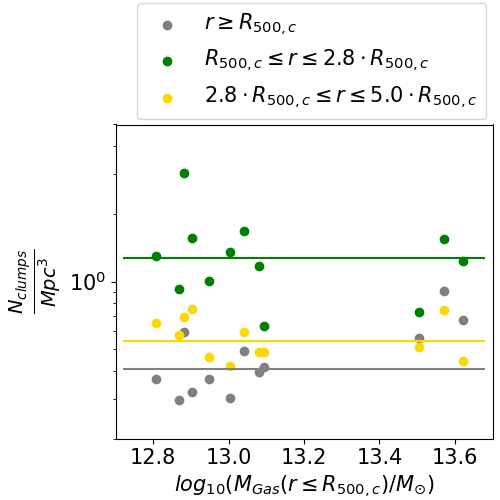}
\caption{Number density of identified clumps per Mpc$^3$ (dots), as a function of the host cluster gas mass. The different colors show different radial selections: gray: $r \geq R_{500,c}$; green: $R_{500,c} \leq r \leq 2.8 \cdot R_{500,c}$; gold: $2.8 \cdot R_{500,c} \leq r \leq 5.0 \cdot R_{500,c}$. The solid lines represent the median values of clumps per Mpc$^3$, computed on the whole cluster sample, using the same color-coding adopted for the dots.}
\label{fig:nclumps_shells}
\end{figure}

We next study the proprieties of clumps as a function of their distance from the cluster center, collecting the clumps into two radial shells.
The inner shell goes from $R_{500,c}$ to $2.8 \cdot R_{500,c}$, while the outer one spans from $2.8 \cdot R_{500,c}$ to $5.0 \cdot R_{500,c}$. As a first step, we analyze the number density of clumps per  Mpc$^{3}$. The results are shown in Fig.~\ref{fig:nclumps_shells}. The median number of identified clumps does not depend significantly on the host cluster's gas mass. However, it decreases with the growing  distance from the cluster center. Moreover, if we  consider the whole volume outside the $R_{500,c}$ sphere, the clump number density slightly increases with cluster's mass. This suggests that massive clusters roughly have the same number of clumps within $R_{500,c}-5.0 \cdot R_{500,c}$, but have more clumps outside $5.0 \cdot R_{500,c}$ compared to less massive ones. Furthermore, we observe that the median clump density values for both the inner and outer shells are greater than what we obtain for the full simulation boxes around the clusters. This indicates, as expected, that clumps tend to be concentrated in the near-vicinities of host clusters.

\begin{figure*}
\includegraphics[width=0.99\textwidth]{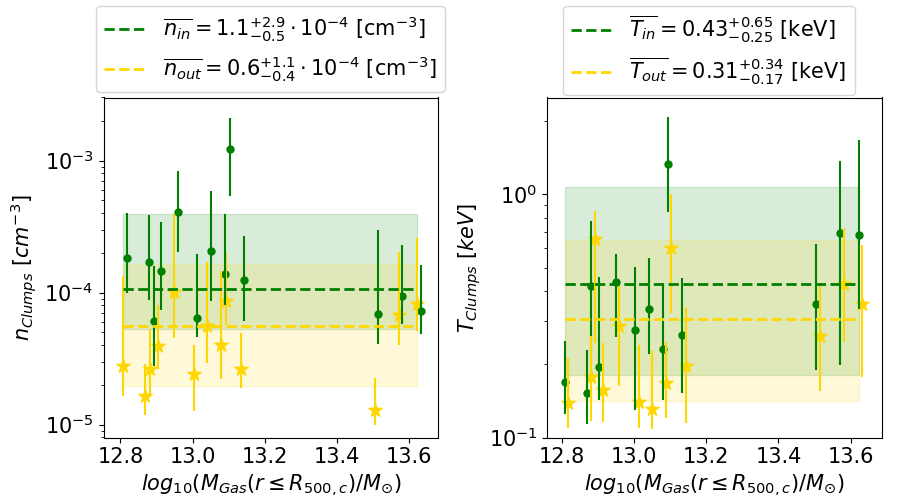}
\caption{Clump density (on the left, in cm$^{-3}$ units) and temperature (on the right, in keV units) as functions of M$_{\rm 500,c}$ cluster gas mass, for the different radial shells. In both the panels, the dots represent the median value, while the error bars are the 16$^{th}$ and 84$^{th}$ distribution percentiles. The colors are the same for both the panels: green represents the inner shell ($R_{500,c} \leq r \leq 2.8 \cdot R_{500,c}$), while gold represents the outer shell ($2.8 \cdot R_{500,c} \leq r \leq 5.0 \cdot R_{500,c}$). The dashed lines represent the median values of density and temperature for the different clump's population (using the same color-code of the dots). The shadow regions are enclosed in the 16$^{th}$ and 84$^{th}$ of the density and temperature distributions.}
\label{fig:n_t_shells}
\end{figure*}

We also investigate possible variation in clump density and temperature with distance from the cluster center. In Fig.~\ref{fig:n_t_shells} we show the results of this analysis. Both density and temperature show a dependence on radial distance from cluster center. We estimate median density and temperature values for the different shells. The clumps in the inner shell are described by a median density of $1.1^{+2.9}_{-0.5}\cdot 10^{-4}$ cm$^{-3}$ and a temperature of $0.43^{+0.65}_{-0.25}$ keV. In the outer shell, we obtain analogous density  $0.6^{+1.1}_{-0.4}\cdot 10^{-4}$ cm$^{-3}$ and temperature $0.31^{+0.34}_{-0.17}$ keV, revealing a decrease of the median  density and temperature with radial distance. By combining the analysis of Fig.~\ref{fig:nclumps_shells} and Fig.~\ref{fig:n_t_shells}, we conclude that clump X-ray detection is expected to be easier in regions closer to a cluster center. 
 
Indeed, in these regions the number of clumps per Mpc$^{3}$ is higher than outer regions and both density and temperature are higher than in the external regions. Higher values of density and temperature imply higher X-ray emissivity. On the other hand, the closer we get to the cluster center, the greater is the X-ray emission due to the cluster itself.

To evaluate the volume and mass contributions of clumps and filaments in different shells, we define a dimensionless function $f$ as
\begin{equation} \label{eq:f}
    f = \frac{\rm m_{k}}{\rm M_{Shell}}  
\end{equation}
where m$\rm _k$ in the numerator represents the (baryon) mass in clumps or filaments, while the denominator is the total baryon mass in the inner or the outer shell, combining the contributions of clumps, filaments and the diffuse gas for the same cluster. We studied also the volume filling factor, $\Phi$,  of the structures, defined as:
\begin{equation} \label{eq:phi}
\Phi = \frac{\rm {\cal V}_{k}}{\rm {\cal V}_{Shell}}.
\end{equation}
As done for the $f$, mass fraction metric, ${\cal V}_{\rm k}$ represents the volume occupied by clumps or filaments, and the denominator is the surveyed volume in each shell.

\begin{figure*}
\includegraphics[width=0.99\textwidth]{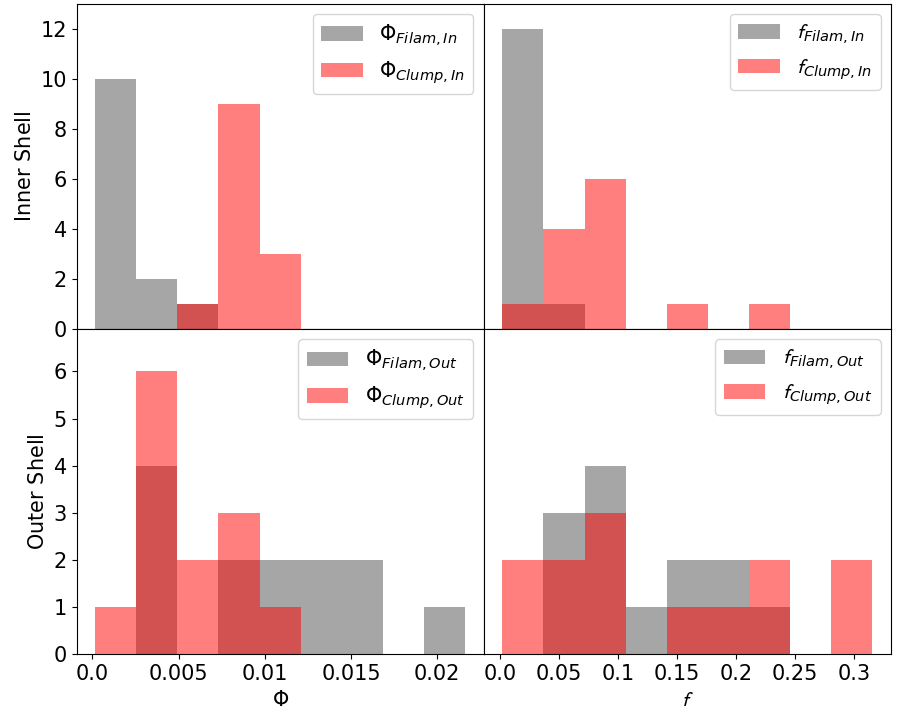}
\caption{Comparison between clump and filament volume filling factor distributions, $\Phi$ and their mass contributions, $f$, for the different adopted shells. In all panels, the red histogram represents the clump distribution, while the gray color identifies the filament distribution. The upper panels show $\Phi$ (on the left) and $f$ (on the right) for the inner shell, while the bottom panels show the same distributions in the outer shell.}
\label{fig:m_v_shells}
\end{figure*}

In Fig.~\ref{fig:m_v_shells} we show the results for both the adopted shells. We observe different trends for different shells. In particular, in the inner shell (upper panels of Fig.~\ref{fig:m_v_shells}) the filament contributions for both $\Phi$ and $f$, are negligible compared with those from the clumps. On the other hand, when we move to the outer shell (bottom panels of Fig.~\ref{fig:m_v_shells}), these behaviors are quite different. The volume filling factor, $\Phi$ of filaments is now twice that for clumps. However, due to the filaments' lower densities, the mass contribution  of filaments, $f$ is comparable to that of clumps. From these results, we conclude that closer to the central cluster only clumps survive interactions with the ICM gas, while in the outer regions both clumps and filaments coexist. Moreover, the detection of filaments is easier far from the cluster's center due to the larger occupied volume. In contrast, the detection of clumps is facilitated in the inner shells thanks to their intrinsic higher densities and temperatures.

 \subsection{Observable X-ray emission proprieties of clumps}

\begin{figure}
\includegraphics[width=0.49\textwidth]{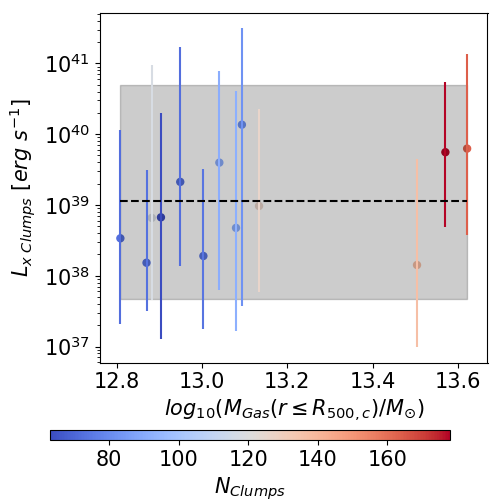}
\caption{Clump soft X-ray emission ([0.3-2.0] keV band) as a function of the central cluster gas mass. The dots represent the median value, while the error bars are the 16$^{th}$ and 84$^{th}$ distribution percentiles. The color-coding identifies the total number of clumps identified by the clump finder algorithm. The black dashed line represents the median value computed for the entire clump population. The gray shadow region identifies the 16$^{th}$ and 84$^{th}$ percentile boundaries of the clump soft X-ray emission distribution.}
\label{fig:l_clumps}
\end{figure}

We investigate the proprieties of the observable X-ray emission of clumps in our simulated sample of clusters. We compute the soft X-ray emissions in the band [0.3-2.0] keV for clumps in the radial range $(R_{500,c}, 5 \cdot R_{500,c})$. We consider the density and temperature in the simulated box and we derive the X-ray emissivity from the sum of the following components:
\begin{equation}
  \rm  Em_i = Em_{ff,i} +Em_{lin,i}= n_{i}^2 \cdot (\Lambda_{ff}(T_{i}) +\Lambda_{lin}(T_{i})) 
\end{equation}

where {\it i} is an index identifying each cell in the box, {\it n} is the gas density in particles per cm$^3$, {\it T} is the gas temperature in Kelvin and  $\Lambda_{ff}$(T$\rm _{i}$) and $\Lambda _{lin}$(T$\rm _{i}$) are respectively the conversion factor for the free-free emission and the lines, at a given temperature. {\b These} were computed as in \citet{va19}, by assuming for simplicity a single temperature and a single (constant) composition for every cell in the simulation using the APEC emission model {\footnote{https://heasarc.gsfc.nasa.gov/xanadu/xspec/manual/Models.html}}. 
We adopt a constant metallicity across all cluster volumes, $Z/Z_{\odot}=0.3$, where $Z_{\odot}$ is the solar abundance in \cite{ag89}.
Then, we consider as X-ray emission from the clump the sum of the contribution of all cells belonging to the identified clump.
In Fig.~\ref{fig:l_clumps}, we show clump X-ray emission as a function of the central cluster mass. Similar to what we found for clump masses and volumes, the X-ray emissions do not show a strong correlations with the  total mass of host clusters. The median emission of our sample of clumps is found to be $0.11^{+4.83}_{-0.10}\cdot 10^{40}$ erg s$^{-1}$. 

\begin{figure*}
\includegraphics[width=0.99\textwidth]{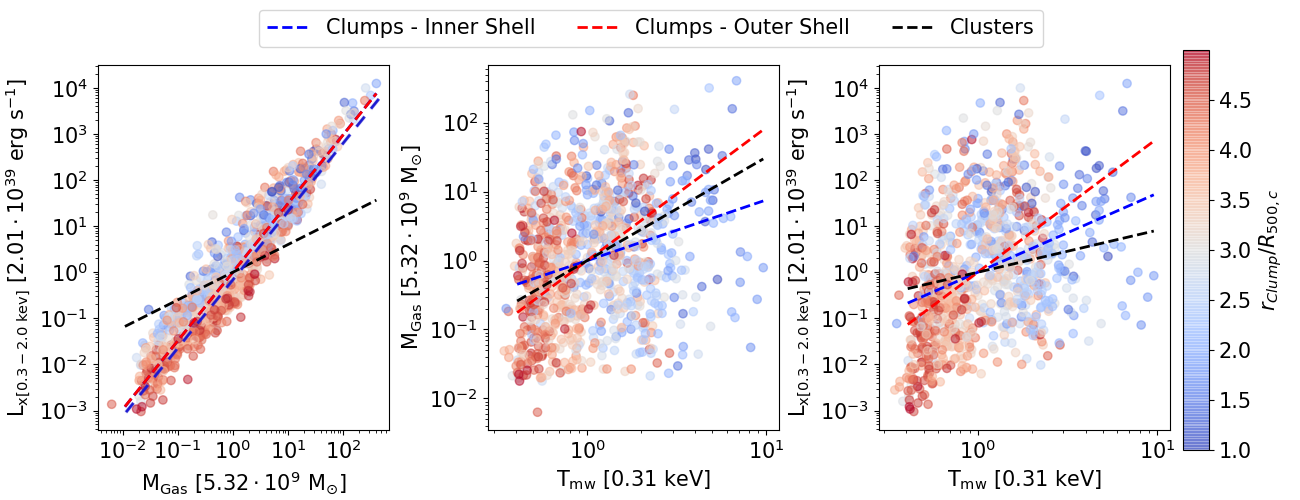}
\caption{Scaling relations L$_{\rm x}$-M$_{\rm Gas}$, M$_{\rm Gas}$-T$_{\rm mw}$ and L$_{\rm x}$-T$_{\rm mw}$, from left to right panel, respectively. The dashed lines identify different selections: clumps selected in the inner region in blue (from R$_{500,c}$ to 2.8 $\cdot$R$_{500,c}$), clumps selected in the outer region in red (from 2.8 $\cdot$R$_{500,c}$ to 5.0 $\cdot$R$_{500,c}$), cluster sample in black. The colorful dots represent single clump. They are color-coded as a function of the clump's radial distance from the cluster center, normalized to  R$_{500,c}$. We note that in the left panel the blue dashed line is shifted arbitrarily for visual purpose. Indeed, it is perfectly over-posted on the red one, due to their identical slope.}
\label{fig:scalerel}
\end{figure*}

We study three different scaling relations for the sample of clumps: X-ray luminosity vs mass (L$_{\rm x}$-M$_{\rm Gas}$), X-ray luminosity vs gas (mass-weighted) temperature (L$_{\rm x}$-T$_{\rm mw}$), and mass vs gas (mass-weighted) temperature (M$_{\rm Gas}$-T$_{\rm mw}$), as shown in Fig.~\ref{fig:scalerel}. We also perform the same analysis on the full clusters in the sample. In that case, we compute the same quantities (L$_{\rm x}$, M$_{\rm Gas}$ and T$_{\rm mw}$) within R$_{500,c}$. Dots in Fig.~\ref{fig:scalerel} are color coded according to clump radial distance from cluster centers. Radial trends are apparent for all quantities displayed. The fitted trends are shown with the dashed lines with proprieties listed in Table~\ref{tab:scalerel}. To prescribe such scaling relations, we use linear fitting functions. For the clump population we sample the range of mass and temperature with 20 equally spaced logarithmic bins. Comparing the slopes obtained for the cluster sample and those proposed by \citet{Lovisari15}, we notice that our results are in good agreement with the slopes expected for simulations. However, they do not match observations. These conflicting behaviors are expected because observed clusters are influenced by radiative energy loss processes that are not included in our non-radiative simulations. 

\begin{table}
\begin{center}
    \begin{tabular}{ c|c|c }
    \bf{Clusters} & \multicolumn{2}{c}{\bf{Clumps}} \\ \hline \hline
     & \bf{1.0$\leq$r/R$_{500,c} \leq$2.8} & \bf{2.8$\leq$r/R$_{500,c} \leq$5.0} \\ \hline
     L$_{\rm x}\propto$ M$_{\rm Gas}^{0.59}$ & L$_{\rm x}\propto$ M$_{\rm Gas}^{1.49}$ & L$_{\rm x}\propto$ M$_{\rm Gas}^{1.49}$\\ \hline
     M$_{\rm Gas} \propto$ T$_{\rm mw}^{1.49}$ & M$_{\rm Gas} \propto$ T$_{\rm mw}^{0.88}$ & M$_{\rm Gas} \propto$ T$_{\rm mw}^{1.93}$ \\ \hline
     L$_{\rm x}\propto$ T$_{\rm mw}^{0.91}$ & L$_{\rm x}\propto$ T$_{\rm mw}^{1.71}$ & L$_{\rm x}\propto$ T$_{\rm mw}^{2.89}$ \\  
    \end{tabular}
    \caption{Scaling relations (M$_{\rm Gas}$-L$_{\rm x}$), (M$_{\rm Gas}$-T$_{\rm mw}$) and (L$_{\rm x}$-T$_{\rm mw}$) for the entire cluster sample and the two clumps subsamples obtained by the inner and the outer shells.}
    \label{tab:scalerel}
\end{center}
\end{table}
To obtain a physical interpretation of these scaling relations, we  compare the computed relations with similar ones measured  by \citet{Eckmiller11} and \citet{Lovisari15} in samples of X-ray galaxy groups. \citet{Eckmiller11} used Chandra observations of 26 objects, and combined those results with results derived from the HIFLUGCS clusters sample. They found that physical differences appear when groups and clusters are studied separately, but that these differences do not affect the proprieties of the scaling relations. Specifically, they found L$_{x}\propto$M$_{500}^{1.34}$, L$_{x}\propto$T$^{2.25}$ and M$_{500}\propto$T$^{1.68}$ for their group sample. \citet{Lovisari15} used a sample of 20 groups observed with the XMM-Newton telescope, and reported L$_{x}\propto$M$_{500}^{1.5}$, L$_{x}\propto$T$^{2.5}$ and M$_{500}\propto$T$^{1.65}$, with a small dependence on the different fitting procedures adopted. 

Comparing our results from  our entire sample of clumps with those obtained observationally by \citet{Eckmiller11} and \citet{Lovisari15}, we conclude that only the M$_{\rm Gas}$-L$_{\rm x}$ relation seems to be in agreement, while the observed L$_{\rm x}$-T$_{\rm mw}$ and M$_{\rm Gas}$-T$_{\rm mw}$ are significantly shallower than our simulation-based results. Nevertheless, we point out that the cluster sample mass ranges of the work by both \citet{Eckmiller11} and \citet{Lovisari15} are different from our clump mass range. However, to further probe the validity of our scaling relations for clumps in a more massive cluster atmosphere, future dedicated X-ray observations will be necessary. 

We also looked for any dependencies of the above scaling relations on distance from cluster center. Clearly, only the M$_{\rm Gas}$-L$_{\rm x}$ relation is unchanged from small to large radii. Both L$_{\rm x}$-T$_{\rm mw}$ and M$_{\rm Gas}$-T$_{\rm mw}$ relations are steeper in the outer analyzed regions, in line with \citet{Eckmiller11} and \citet{Lovisari15}. This suggests that outside $\sim$3$\cdot$R$_{500,c}$ clumps have not interacted with the central cluster and their physical proprieties follow the self-similar assumption, so they can be considered as mass-rescaled galaxy clusters. Recently, using data from THE THREE HUNDRED project, a collection of 324 simulated galaxy clusters, \citet{Mostoghiu21} conclude that infalling sub-haloes lose their gas quicker when they are closer than $\sim$1.5$\cdot$R$_{200}$. This suggests that closer than $\sim$2.5$\cdot$R$_{500}$ the clump-cluster gas interactions could strongly change the physical proprieties of clumps and consequently the estimation of the scale relations.   
From Fig.~\ref{fig:scalerel}, we also note that clumps are less massive, colder and less luminous with increasing radial distance from the cluster center. All these trends influence their chance of detection around galaxy clusters. By combining the trends shown in Fig.~\ref{fig:scalerel} and the results of Sect.~\ref{sec:radialtrends}, the shell just outside $R_{500,c}$ seems to be the region that might ensure the best conditions for the detection of gas clumps using X-rays. Nevertheless, the study of clumps beyond $\sim$2.5$\cdot$R$_{500}$ enables reconstruction of physical proprieties which are not affected by interaction with the central galaxy cluster's gas.

\section{Conclusions} \label{sec:conclusions}

The presence of clumps and filaments in the proximity of galaxy clusters, and physically connected to the gas atmosphere of the host halo 
was predicted early on by cosmological simulations and more recently supported by X-ray observations \citep{eckert15,simionescu17,reiprich20}. However, due to the low X-ray emissivity of such structures, a complete physical picture of these structures supported by observations is still missing. 
In this work, we use a catalog of 13 galaxy clusters simulated at uniformly high resolution with the cosmological code {\enzo} \citep[see Sect.~\ref{sec:ITASCA}; ][]{va17turbo,wi17b,2018MNRAS.481L.120V} to introduce and test novel strategies that could be used to identify and characterize such clumps and filaments both from simulations, and through observations.

We have verified (see Appendix A) that the spatial resolution used in our simulations is suitable to capture and characterize clumpy and filamentary accretions formed in these non-radiative simulations, in a  reasonably converged way.
We expect, however, that some of the statistics extracted from our sample (especially the thermodynamical proprieties of clumps) may change after the adoption of more sophisticated treatments of  chemistry, cooling and galaxy-formation related processes.

Specifically, we have introduced two different algorithms to robustly identify, and distinguish between, clumps and filaments in each of our simulated cluster volumes.
Our clump finder algorithm is based on work by \citet{2013MNRAS.428.3274Z} and \citet{2013MNRAS.432.3030R}, where clumps were classified as a tail of the log-normal mass density probability distribution function, corresponding to the densest 1\% of cells within each radial shell around a cluster center. 
In order to associate cells in our simulations with individual clumps,
we implement a clustering procedure for neighboring selected cells, based on the baryon mass within a chosen enclosure radius (Sect.~\ref{sec:clump}).

Owing to their more complex and extended morphologies, filaments are more challenging to define than clumps, both observationally and numerically. 
While clumps can be robustly identified by local overdensity conditions, filaments require a more sophisticated approach \citep[see][for a detailed reviews on numerical definitions of filaments]{Libeskind18}, especially when they overlap with the cluster ICM.
We have, thus, developed a new proxy tool targeting the parameters of radial velocity and entropy associated with gas undergoing accretion from cosmic web filaments onto our simulated galaxy clusters (as in Eq.~\ref{eq:filam}). 

We stress that detailed physical analysis of clumps and filaments in our work is limited by the restricted range of physical processes included in our (non radiative) suite of cluster formation simulations. 

\citet{Galarraga20} recently applied simulations with a richer set of physical processes from the Illustris-TNG suite \citep{Nelson19}. The phase diagrams of gas in their simulations show proprieties distinct from our work (Fig.~\ref{fig:IT90_2_phase}), supporting the notion that non-gravitational physical processes related to galaxy formation, as well as radiative gas cooling, can significantly affect the evolution of gas thermodynamics. In their work,  those authors divided the gas component into five different phases, based on threshold values of  density and temperature. We note that in our simulations their "halo gas" phase is completely missing, while the warm circumgalactic medium (WCGM) is significantly reduced. We would, indeed, expect those phases to depend on details of gas thermodynamics. In our work we focus on simpler and numerically less expensive non-radiative simulations, thus assessing the impact of the processes described above on the observational proprieties of filaments that it will be a mandatory step to consider. 
We can refer to an earlier inspection of the effects of radiative gas physics and AGN feedback on the distribution of density fluctuations in (similar) ENZO simulations of galaxy clusters  \citep{vazza13}. This analysis found, on the one hand, that the physical proprieties of clumps are modified by details of baryon physics, but also that the large-scale organization of clumps and filaments around them is not significantly modified by this. Concerning the effect of spatial resolution, our simulations are robust in the sense that they adopt a fix spatial resolution in the cluster region, rather than adaptive mesh refinement. Additional resolution tests presented in Appendix~\ref{app:testres} also show that the impact of spatial/force resolution on the distribution of the gas clumping factor and on the formation of single clumps is small once a resolution $\leq 20 \rm ~kpc$ is considered throughout their development.

The combination of the above results suggests that the suite  of simulations we used for this work can provide a robust initial characterization of the morphology, distribution and large-scale correlation of clumps and filaments, which are largely dominated by purely gravitational processes. However, the detailed thermodynamical proprieties of clumps (and, to a lesser extent, also of filaments) are expected to depend on specific prescriptions for baryonic physics. In this respect, our non-radiative simulation setup can only be considered as a toy model to study the feasibility of future observational investigations. Incidentally, we note that early tests presented in \citet{vazza13} compared the average number of clumps detected in high resolution non-radiative cosmological simulations (similar to the ones analyzed in this work) and the catalog of point-like sources identified by the X-ray analysis of ROSAT exposures of A2142, finding an agreement within a factor $\sim 2$ across the cluster volume (see their Fig. 11).

Given the above caveats, our analysis (enabled by our new detection algorithms) of clusters extracted from the Itasca simulation set can be so summarized: 
\begin{itemize}
    \item The density and temperature of clumps are independent of the mass of the cluster where they reside. The distributions have median values $0.6^{+1.6}_{-0.4}\cdot 10^{-4}$ cm$^{-3}$ and $0.3^{+0.4}_{-0.2}$ keV, respectively. The clumps identified  in the simulations have a median mass   $0.44^{+3.53}_{-0.38}\cdot 10^{10}$ M$_{\odot}$ and a median volume $0.4^{+1.6}_{-0.3}\cdot 10^{-2}$ Mpc$^{3}$, independent of the host cluster mass. 
    \item We examined density and temperature dependencies for both clumps and filaments as functions of the host cluster's mass, finding no strong correlation between filament density and cluster mass. On the other hand, we detected a small positive correlation between the filament temperature and the host cluster mass. The median density and temperature of our sample of filaments have values  $1.3^{+3.4}_{-0.5}\cdot 10^{-5}$ cm$^{-3}$ and $0.04^{+0.05}_{-0.02}$ keV, respectively, in both cases these are lower than the median values of clumps. 
    \item By comparing the density and temperature distributions of clumps and filaments populations in each cluster of our sample, we investigated the possible correlations between these quantities. 
    We use the Pearson's correlation index to describe these correlations and we obtain $\rm \rho_{XY}$=0.66 and $\rm \rho_{XY}$=0.54, for density and temperature respectively. 
    \item When the simulated volume of each cluster is divided into a pair of radial shells, we observe a significant drop (a factor two or more) in the number of clumps going from the inner ($ \rm 1.0 \le r/R_{500,c} \le 2.8$) to the outer shell ($ \rm 2.8 \le r/R_{500,c} \le 5.0$). 
    We study the variations in clump density and temperature with distance from cluster center. In both, we also report a decrease (by a factor $\sim$2) in the typical density and temperature of clumps as a function of distance from the cluster center, as well as in their mass and volume filling fraction.   
    We observe how in the inner shell the contribution of filaments, both in mass and volume, are less than for the clumps. On the other hand, in the outer shell, the volume occupied by filaments is on average $\sim$2 times the one occupied by clumps. 
    \item Lastly, we study the scaling relations M$_{\rm Gas}$-L$_{\rm x}$ M$_{\rm Gas}$-T$_{\rm mw}$ and L$_{\rm x}$-T$_{\rm mw}$ and their dependence with radius. 
    We compare the behaviors of the computed clumps scaling relations with what has been obtained previously for structures with larger masses, like galaxy groups \citep{Eckmiller11,Lovisari15}. 
    We note that only for $r \la 2.5 \cdot R_{500,c}$ are these relations for clumps in agreement with ICM clump relations proposed by \citet{Eckmiller11} and \citet{Lovisari15}. This suggests that close to the cluster center, the physical proprieties of infalling clumps are strongly affected by the dynamical interactions between the clumps and the intracluster medium, whereas in the outer regions clumps keep their identity and behave more like self-similar cosmological formation expectations. Moreover, we observe that clumps are hotter, more massive and more luminous if identified closer to the cluster center. This result (albeit, perhaps limited to non-radiative simulations) is a useful guide for the thermodynamic analysis of X-ray detectable clumps associated with clusters.   
\end{itemize}

Our work focuses, both, on global and local proprieties of gas clumps and filaments surrounding galaxy clusters. 
We expect, however, that some of the statistics (especially concerning the innermost thermodynamical proprieties of clumps) presented in this paper may change when more sophisticated treatment of chemistry, cooling and galaxy-formation related processes are considered. 

In summary, we consider the measured statistical association between  clumps and filaments a first, promising step to assess the overall evolution of matter accretions in galaxy clusters with future X-ray observations. 
Clumps are already on the verge of being detectable with long, soft X-ray exposures, and surely will be in the near future using the next generation of X-ray telescopes (e.g., Athena\footnote{\url{https://www.the-athena-x-ray-observatory.eu/}}, AXIS\footnote{\url{https://axis.astro.umd.edu/}}, Lynx\footnote{\url{https://www.lynxobservatory.com/}}).
Our simulations suggest that the
region of clusters (e.g., between $R_{500,c}$ and $3 \cdot R_{500,c}$) may represent a sweet spot to detect and study the thermodynamics of clumps and of cosmic baryons, prior to strong interaction with the ICM. On the other hand, in order to study the physical proprieties of filaments, even more peripheral regions ($\gtrsim 3 \cdot R_{500,c}$) should be targeted by observations. We postpone to a forthcoming paper a detailed study on how the clumps and filaments analyzed in the present work can be detected and characterized with dedicated exposures with the next-generation of X-ray instruments (X-IFU and WFI onboard Athena, in particular).

\section*{Acknowledgements}
The cosmological simulations described in this work were performed using the {\enzo} code (http://enzo-project.org), which is the product of a collaborative effort of scientists at many universities and national laboratories. S.E. acknowledges financial contribution from the contracts ASI-INAF Athena 2019-27-HH.0,
``Attivit\`a di Studio per la comunit\`a scientifica di Astrofisica delle Alte Energie e Fisica Astroparticellare''
(Accordo Attuativo ASI-INAF n. 2017-14-H.0), INAF mainstream project 1.05.01.86.10, and
funding from the European Union’s Horizon 2020 Programme under the AHEAD2020 project (grant agreement n. 871158).
F.V. acknowledges financial support from the European Union's Horizon 2020 program under the ERC Starting Grant "MAGCOW", no. 714196. 
T.J. acknowledges financial support from the US NSF through grant AST1714205.
The PhD grant supporting M.A. is co-funded from INAF and ERC Starting Grant "MAGCOW", no. 714196.

\bibliographystyle{aa}
\bibliography{biblio}

\appendix

\section{Numerical resolution effects on clumping factor and clump number density} \label{app:testres}
 
\begin{figure}
\begin{center}
\includegraphics[width=0.225\textwidth]{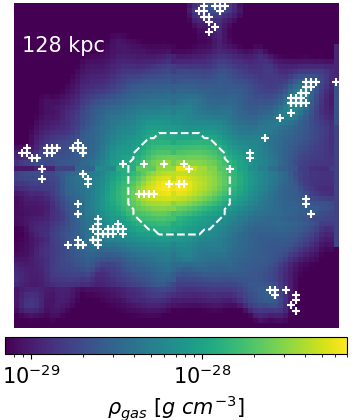}
\hspace{4.1pt}
\includegraphics[width=0.232\textwidth]{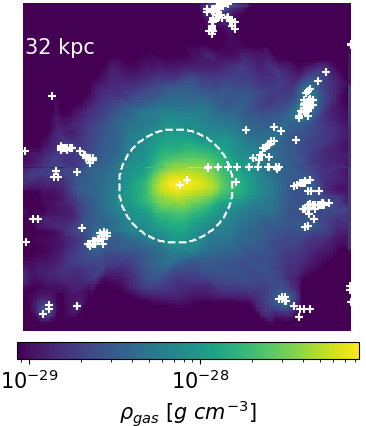}
\hspace{8.6pt}
\includegraphics[width=0.239\textwidth]{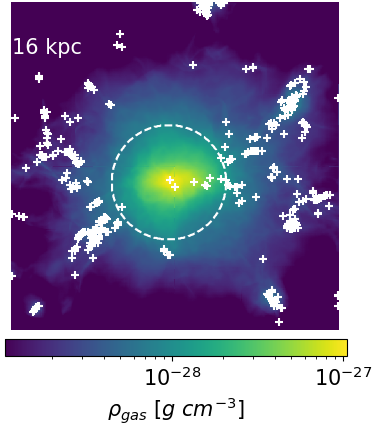}
\hspace{1.0pt}
\includegraphics[width=0.237\textwidth]{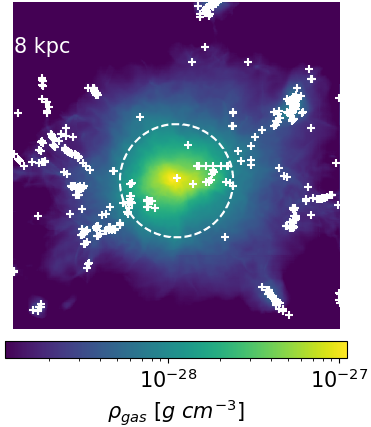}
\end{center}
\caption{Projected gas density for the test cluster, at four different level of numerical resolution (top left: 128 kpc; top right: 32 kpc; bottom left: 16 kpc; bottom right: 8 kpc). The dashed circle represent R$_{\rm 500,c}$, while the white crosses are the center of the identified clumps.}
\label{fig:res_map}
\end{figure}

\begin{figure}
\includegraphics[width=0.49\textwidth]{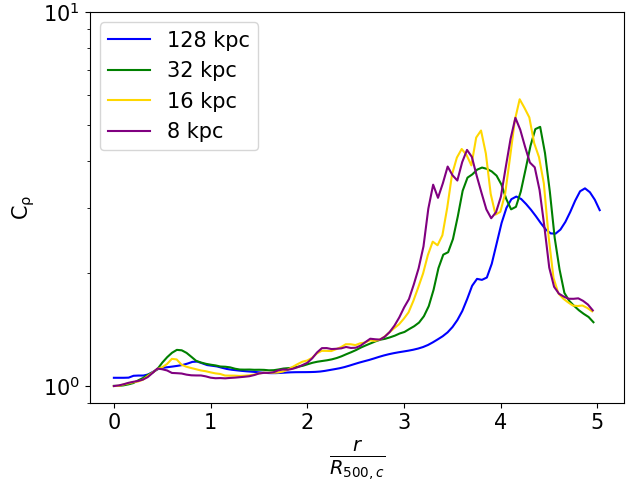}
\caption{Clumping factor for the test cluster at different levels of numerical resolution.}
\label{fig:clumpingfactor_res}
\end{figure}

We tested the effect of the spatial and force resolution on the measure of the
clumping factor and on the number of identified clumps, to understand how much are the findings of our main paper restricted to the particular adopted resolution. To this end, we analyzed four different resimulations of the same
cluster of M$_{500,c}$=8.5$\cdot  10^{14}$ M$_{\odot}$ and R$_{500,c}$=1.4 Mpc, obtained for an increasing level of maximum adaptive mesh refinement, as in \citet{2018MNRAS.474.1672V}. The adaptive mesh refinement follows the local gas overdensity, and the maximum resolution goes from 128 kpc in the lowest resolution case, to 8 kpc in the highest resolution run adopted here (see Fig.~\ref{fig:res_map} for the related density maps and detected clumps). 
We followed the same procedures of the main paper, and computed the radial profiles of gas clumping factor and clumps as a function of resolution (Fig.~\ref{fig:clumpingfactor_res}). 

It can be noticed that in the central regions (up to $\sim$2.5$\cdot R_{500,c}$), the clumping factor is weakly affected by changing of numerical resolution. However, in the external regions, the differences introduced by the AMR levels account for less than 30\% of the clumping factor at any radii. This trend ensures that the evaluation of clumping factor is weakly affected by the simulations' numerical resolution - at least for the non-radiative gas physics used in this work. 

\begin{figure}
\includegraphics[width=0.49\textwidth]{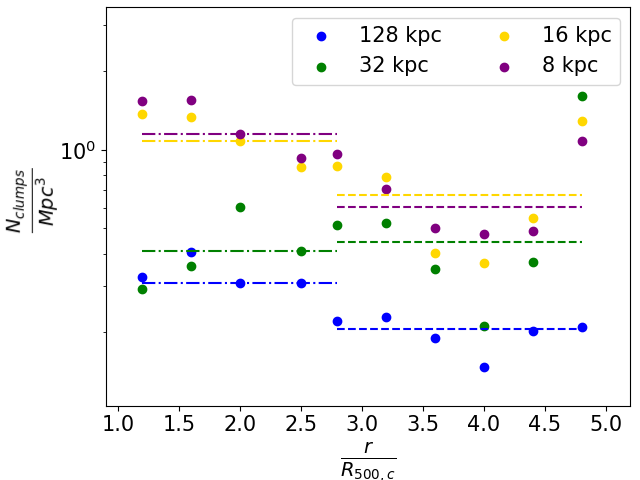}
\caption{Clumps' number density for the same cluster at different levels of numerical resolution in function of radial distance from the cluster's center. The dot-dashed lines represent the median value computed between R$_{500,c}$ and 2.5$\cdot$R$_{500,c}$, while the dashed ones represent the median value computed from 2.5$\cdot$R$_{500,c}$ to 5$\cdot$R$_{500,c}$. The color-coding of these lines is the same adopted for the dots.}
\label{fig:numberdensity_res}
\end{figure}

Second, we tested the effects of numerical resolution on the clumps number density. Using the same AMR level adopted for the clumping factor, we use ten equally spaced radial shell from R$_{500,c}$ up to 5$\cdot$R$_{500,c}$. From Fig.~\ref{fig:numberdensity_res}, we notice that increasing the numerical resolution, the clumps number density increases. In particular, we observe a difference of factor $\sim$3 between the lower numerical resolution and the higher one. However, we also notice that from 16 kpc to 8 kpc, the median values computed in the inner regions (from center to 2.5$\cdot$R$_{500,c}$) and the outer ones (from 2.5$\cdot$R$_{500,c}$ and 5$\cdot$R$_{500,c}$) are quite constant. The observed trend is exactly as in them main paper, and the relatively little evolution with resolution suggests that spatial resolution of the cluster sample used in the main paper ($\sim$19.6 comoving kpc) is suitable to study clumps and clumping free from numerical artifacts. 

\section{Summary table of Pearson's correlation indexes and their significance} \label{app:pearson}

\begin{table*}
\begin{center}
    \begin{tabular}{ c|c|c|c }
    \bf{Proprieties} & \bf{Pearson's correlation index} & \bf{t$^{*}$} & \bf{P-value} \\ [1ex] \hline \hline 
    Filaments' temperature vs. cluster's mass & 0.88 & 6.15 & $\le$0.001 \\ \hline
    Density clumps vs. filaments & 0.66 & 2.91 & $\le$0.02 \\  \hline
    Temperature clumps vs. filaments & 0.54 & 2.13 & $\le$0.1 \\ \hline
    \end{tabular}
    \caption{Pearson's correlation index, test statistic t$^{*}$ and P-value for the different correlations analyzed in Sect.~\ref{sec:physicalclumpfilam} }
    \label{tab:pearson}
\end{center}
\end{table*}

The Tab.~\ref{tab:pearson} summarize the statistical indices adopted for the analysis of the correlations between different clumps and filaments proprieties used in Sect.~\ref{sec:physicalclumpfilam}.
We compute Pearson's correlation index as:
\begin{equation}
    \rm \rho_{XY} = \frac{\sigma_{XY}}{\sigma_{X}\cdot \sigma_{Y}}
\end{equation}
where $\rm \sigma_{XY}$ is the covariance between the quantities X and Y, while $\rm \sigma_{X}$ and $\rm \sigma_{Y}$ are the standard deviations of X and Y. Then, the test statistic t$^{*}$ is given by the following equation:
\begin{equation}
    \rm t^{*} = \frac{\rho_{XY}\cdot \sqrt{n-2}}{\sqrt{1-\rho_{XY}^2}}
\end{equation}
where "$\rm \rho_{XY}$" is Pearson's correlation index and "n" is the total number of clusters in the sample. We compare the computed t$^{*}$ values with the tabulated Student's t distribution for the relative number of freedom degrees (12, N$_{Cluster}$-1) and we derive the P-value for each Pearson's correlation index shown in Tab.~\ref{tab:pearson}. 

As already discussed in Sect.~\ref{sec:physicalclumpfilam}, many of the studied relations show a strong level of correlation, which are also confirmed by the significance given by the P-value. 

\end{document}